\documentclass[aps,amsmath,amsfonts,11pt]{revtex4}
\usepackage{graphicx}
\usepackage{epstopdf}
\usepackage{epsfig,graphicx}
\usepackage[english]{babel}
\usepackage{amsfonts}
\usepackage{amsmath}
\usepackage{latexsym}
\usepackage{graphics,bm}
\usepackage{dcolumn}
\usepackage{natbib}
\usepackage{bm}
\usepackage{rotating}

\begin{document}

	\title{Coherent control of quantum and entanglement dynamics via periodic modulations in optomechanical semi-conductor resonator coupled to quantum-dot excitons}
	
	\author{ Vijay Bhatt$^{1}$, Pradip K. Jha$^{1}$, Aranya B. Bhattacherjee$^{2}$, Souri Banerjee$^{2}$}
	
	\address{$^{1}$Department of Physics, DDU College, University of Delhi, New Delhi 110078, India }
	\address{$^{2}$Department of Physics, Birla Institute of Technology and Science, Pilani, Hyderabad Campus, Hyderabad - 500078, India}

	\begin{abstract}
		We systematically study the influence of simultaneously modulating the input laser intensity and quantum dot (QD) resonance frequecy on the mean-field dynamics, fluctuation energy transfer and entanglement  in a optomechanical semi-conductor resonator embedded with a QD. We show that the modulation and the hybrid system can be engineered to attain the desired mean-field values, control the fluctuation energy transfer and the entanglement between the various degrees of freedom. A remarkably high degree of entanglement can be achieved by modulating only the QD frequency. The interplay between the two modulations leads to an entanglement which lies between that generated solely by modulating either the QD or the pump laser intensity.  A transition from low stationary to large dynamical entanglement occurs as we switch on the modulation. This study opens up new possibilities for optimal control strategies and can be used for data signal transfer and storage in quantum communication platforms.
	\end{abstract}

	\maketitle

	\section{Introduction}
	Theoretical studies and tremendous technological development over the last few decades have allowed a substantial degree of control over matter in a wide range of physical systems, ranging from electrons, photons, atoms to larger solid-state systems such as quantum dots and superconducting circuits. This has opened up the possibility of testing novel quantum mechanics and allowing us, among other things, to take major steps forward in investigating the quantum regime of macroscopic objects. In this context, the controlling of nano- and micromechanical oscillators at the quantum level is one of the key goals in today's research in quantum optomechanics.	Quantum optomechanics \citep{marq,asp,kipp,milb,milb-2,asp-2} (i.e., the interaction of light radiation pressure with mechanical systems) comes as an effective and a well-developed method for such a kind of quantum manipulation. The operation of the radiation pressure can be used to cool a (nano)micromechanical oscillator to its motional ground state\citep{gene}; this is a necessary step for quantum manipulation and could not be accomplished by direct means such as cryogenic cooling. Cooling has been experimentally demonstrated for a variety of physical implementations, including micromirrors in Fabry-Perot cavities \citep{gor,vitali}, microtoroidal cavities \citep{ver}, or optomechanical crystals \citep{chan}. There is a close correlation between quantum optomechanics and nonlinear quantum optics, making it possible to model many (if not all) optomechanical effects based on well-known nonlinear-optical effects. As a result, optomechanics is a simple means of regulating the mechanical resonator at the quantum level.
	
	  In particular, Refs.\citep{mari-1,mari-2,galve} introduced an effective way of enhancing the generation of quantum effects, which relies on applying a periodic modulation to some of the system parameters. A similar result has also been found in the analogous context of nanoresonators and microwave cavities \citep{woo}.	In multiply connected systems,  the optomechanical coupling can be harnessed in the transfer of quantum states from one system to another which has useful applications in the field of quantum communication. Specifically, the state of the cavity field can be coherently transferred, stored, and retrieved from the mirror\citep{pal,saf}. A strong coupling between a two-level system and a mirror in motional degrees of freedom leads to a coherent exchange of average energy between them, which is yet another novel system that can be used in quantum information processing systems \citep{ham,wall}.

	 In view of the above scenario, the study of quantum entanglement plays a crucial role if the field of quantum theory, serving as a fundamental tool in processing of quantum information. Quantum entanglement is the study of non-local correlations between multiple degrees of freedom. Correlations attributed to entanglement is fast emerging as an essential tool in quantum technologies such as quantum computation, sensing and communication \citep{nielsen}. Entanglement has been observed between optical and mechanical resonator modes in a double-cavity system \citep{pinard,pater,bhatta,wipf}. The purpose of entangled optomechanical systems \citep{vita-2,vita-3} is to create quantum communication networks where mechanical modes act as local nodes for the retrieval and storage of quantum information and optical modes for the transfer of information between these nodes \citep{manici,pira-1,pira-2,pira-3}. Such hybrid systems are therefore useful in the fields of entanglement swapping \citep{pira-3},  quantum teleportation \citep{manici,pira-2}, and quantum telecloning \citep{pira-1}. The fast-growing field of cavity optomechanics \citep{aspel-1,meys-1,sharma} provides a better platform to prepare such an entangled state in mechanical motion. In recent times, the research in this area has virtually exploded. Many studies have been reported on the entanglement generation between a cavity field and a mechanical oscillator \citep{abdi,kuyz,hofer,wang,rakh} relying on the coupling of generic radiation-pressure. On the other hand, the modulation-assisted driving of a hybrid optomechanical device gives rise to interesting and rich entanglement dynamics which have been reported \citep{roger,yuan,blatt}. More recently, enhanced entanglement has been reported in a system comprising of a two-way coupled quantum dot (QD) inside a hybrid cavity \citep{li1}. Mari and Eisert \citep{mari-1} demonstrated that using an appropriate time-periodic driving, entanglement between a cavity optical mode and a mechanical mode can be enhanced. Simultaneous application of more than one modulations can lead to suppression or enhancement of quantum effects.
	 
	 Quantum optics of semiconductor  micro-cavity deals with light-matter interaction inside meso/nanoscale semiconductor structures. These hybrid systems are fast emerging as a robust and scalable platforms for implementing state of the art technologies for a wide variety of quantum optical devices suitable for quantum communications and quantum networks \citep{mabuchi, sri, kimble}, due to their high tunability and large density of states  \citep{vah1, vah2, khit, tang, ali, aranya}. Particularly, photonic crystal based micro cavities have opened new research frontiers due to their ultra high quality factor and highly confined and extremely small mode volume. The remarkable advantage of solid state based devices is that they can be easily fabricated and integrated  into  large  scale arrays for quantum networks and quantum information  processing applications. 
	
	In view of the interesting developments that have taken place in semiconductor and optomechanical based systems over the past few years in the direction of quantum manipulation of hybrid systems for practical applications, we analyze in this paper the dynamics of a hybrid optomechanical solid-state system in which we simultaneously modulate the QD resonance and the amplitude of the pump laser laser which drives the cavity. Although quantum optomechanics is now widely studied, the quantum effects in simultaneously modulating the quantum dot (QD) frequency and amplitude of the pump laser has not been studied earlier. We show that the coherent exchange of energy fluctuations between the various degrees of freedom of the hybrid system can be controlled by this simultaneous modulation. We also demonstrate that the simultaneous modulation of two system parameters can enhance the coherent control over the quantum entanglement between the various degrees of freedom. We study the mean-field dynamics, quantum fluctuations and quantum entanglement and explore possibilities in the direction of developing optimal quantum control protocols for time modulated solid-state based systems. An interesting question that arises is how simultaneous time modulation of two parameters can be exploited to control specific quantum properties of the system that can be used in some applications.  Immediate results of our research can be used in quantum key distribution (QKD), where quantum entanglement can be utilized to quantify knowledge gained by an adversary \citep{ekert} and to enable encryption \citep{acin}. Long term applications include research directed towards the development of a quantum internet \citep{kimble,wehner}. Such a network includes quantum nodes that are capable of producing, storing, detecting, or verifying quantum entanglement and mesoscopic  optomechanical solid-state systems are promising candidates for such quantum nodes. The experimental  aspects of periodically modulating the QD resonance and pump laser amplitude has already been explored in the context of optical switching \citep{kabuss,fara,arka-1,lei}. A single QD coupled with an optical cavity can modulate the optical signal with very low control power since the system's active volume is very low. The QD resonance is controlled via the Quantum Confined Stark Effect (QCSE) by electric signal. The modulation of the QD resonance  controls laser transmission through the cavity. A lateral electric field to shift the QD resonance has been used\citep{fara}, thereby modulating the light \citep{arka-1}.

	The organization of the paper is as follows. In section 2, we discuss the theory and model of the system and derive the quantum Langevin equations. In section 3, we follow  analytical and numerical techniques to explore the mean-field dynamics of the system. Using perturbative means, we derive analytical solutions for the mean values and show that the system acquires the same period of modulation and approaches the limit cycle in the long time limit. The numerical results agree well with the analytical results.  Further we derive the quantum Langevin equations for the corresponding quadrature fluctuations and study the dynamics of fluctuations energy transfer between the various degrees of freedom of the system. In section 4, we discuss in detail the entanglement dynamics of the system and compare the no-modulation, one-modulation and two-modulations results. We look into the interesting aspect that how the two modulations alter the  quantum dynamics in comparison to the one-modulation picture. Finally in the last section we present the conclusions.

	\section{Theory and model}
	In this article, the model considered is shown in Fig.1, where a single optical mode is interacting with a single mechanical mode in the presence of a QD whose frequency is modulated. The cavity is driven by an external pump laser as shown, whose amplitude is also modulated. In this paper, the motivation of taking DBR (Distributed Bragg Reflector) based optomechanical system is to overcome two obstacles. The first one is to integrate two mirrors on a chip and the second is to involve a moving part within \citep{Hu}. One DBR is fixed and the other acts as the mechanical flexible part, as shown in Fig.1. The confinement of light along the x-direction and the y-z plane is provided by DBR and air-guiding dielectric respectively. There are known techniques by which the confinement of light along the longitudinal and transverse direction in the DBR can be achieved \citep{thon,sonam,arxiv-1,madhav,vijji-2}. DBR mirror comprises of layers of the low and high refractive index of quarter-wavelength having AlGaAs as the first and last layer. The refractive index of AlGaAs is greater than that of air and lower than that of GaAs \citep{choy}. Hence, this structure results in a high-quality-factor semiconductor optical cavity. The total Hamiltonian for the system in the rotating frame of pump laser can be written as,
	
	\begin{figure}[h]
		\includegraphics [scale=0.4]{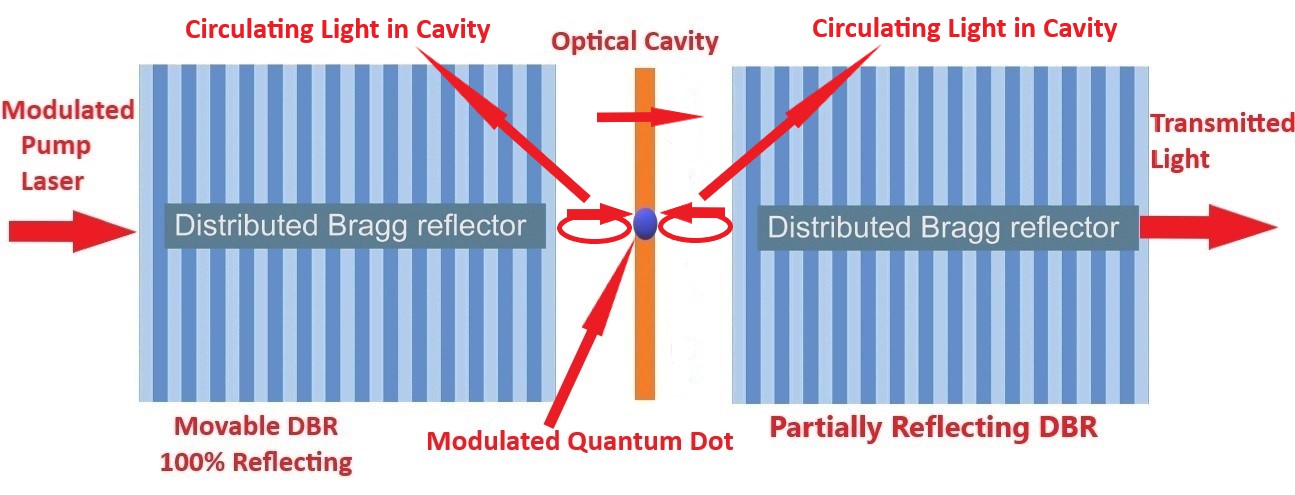}\\
		\caption{(Color online) Schematic display of the configuration discussed in the text. DBR mirrors with the cavity is shown in figure. The left side DBR is movable and quantum dot is placed in between the cavity. The intensity of the pump laser is modulated using known techniques like electro-optic Q switching. The QD resonance is controlled by electrical signal via quantum confined stark effect.}
	\end{figure}
	
	\begin{eqnarray}
	\hat{H}&=& \hbar\Delta_{c}a^{\dagger}a +\frac{p^{2}}{2m} + \frac{1}{2}m\omega_{m}^{2}q^{2}+ \hbar\Delta_{d}(t)\sigma_{+}\sigma_{-} - \hbar G a^{\dagger}a q + \hbar g_{0}(\sigma_{+}a + \sigma_{-}a^{\dagger}) \nonumber \\
	&+& i\hbar E(t)\left(a^{\dagger}-a\right).
	\end{eqnarray}   
	
	Here, $\Delta_{c}= \omega_{c} - \omega_{L}$, $\Delta_{d}= \omega_{d}-\omega_{L}$, $\Delta_{d}(t)=\frac{\Delta_{0}[1-Cos(\omega_{e}t)]}{2}$   and  $E(t)= E_{0} + \epsilon cos(\Omega t)$. In the above Hamiltonian (Eqn.1), $a^{\dagger}(a)$ is the creation (annihilation) operator for the optical cavity field. The dimensionless momentum and position operators of the oscillator is defined by $p$ and $q$ respectively and these satisfy $ [q,p]$ $=$ i$\hbar$. Also, $\Delta_{d}(t)$ is the modulated detuning of the QD, where $\Delta_{0}$ is the maximum detuning of the QD resonance (this is proportional to the amplitude of the electrical signal applied to tune the QD) and $\omega_{e}$ is the frequency of the modulating electrical signal. Here $G$ and $g_{0}$ is the optomechanical coupling constant and coupling between the optical cavity and quantum dot respectively. E(t) is the external drive intensity having frequency $\Omega$, $E_{0}$ is the constant driving amplitude that takes real value and $\epsilon$ is the amplitude of the modulation. 
	
	To study the dynamics of the given system, we write the quantum Langevin equations for the operators from Hamiltonian (Eqn.1) as,
	
	\begin{equation} 	
	\dot{a}=-i\Delta_{c}a-\kappa_{a}a+iGaq + E(t)- i g_{0}\sigma_{-} + \sqrt{2\kappa_{a}}a^{in},
	\end{equation} 
	\begin{equation} 	
	\dot{p}=-\omega_{m} q+ Ga^{\dagger}a-\gamma_{m} p+ \zeta(t),
	\end{equation}
	\begin{equation}
	\dot{q}=\omega_{m}p,
	\end{equation}
	\begin{equation}
	\dot{\sigma_{-}}=i\Delta_{d}(t)\sigma_{z}\sigma_{-} + ig_{0}a\sigma_{z}-\kappa_{d}\sigma_{-}.
	\end{equation}
	
	Here $\gamma_{m}$ and $\kappa_{a}$ are the mechanical and optical damping rates, respectively. Radiation vacuum input noise is represented by $a^{in}$ obeying the standard correlation relation, $\left<a^{in\dagger}(t)a^{in}(t)+a^{in}(t')a^{in\dagger}(t)\right>= \delta(t-t')$ and $\frac{1}{2}\left<\zeta(t)\zeta(t')+\zeta(t')\zeta(t)\right>=\gamma_{m}(2n_{b}+1)\delta(t-t')$ denotes the Brownian noise operator. Here, $n_{b}=1/exp(\hbar \omega/k_{B}T-1)$  is the mean phonon number of the mechanical bath which gauges the temperature T of the system \citep{gio,liu,xu}.
	
	To solve equation(2)-(5), we use mean-field approximation to express relevant operators as sum of the mean values (large) and fluctuation terms (small), i.e., $O= \left <O (t) \right >+\delta O$  =$(O= q,p,a,\sigma)$ are the quantum fluctuation operators with zero-mean around c-number mean values $\left < O(t) \right >$. The standard linearization techniques can be applied to the Eqns.(2)-(5), when we assume that $|\left < O(t)\right >| >> 1$ (strong coherent driving regime).  In this way we can write Heisenberg equation for the mean values $\left < O(t) \right >$  in the form of a set of classical nonlinear differential equations
	
	\begin{equation}
	\left <\dot{q} \right >=\omega_{m} \left <p \right >,
	\end{equation}
	\begin{equation} 	
	\left < \dot{p} \right >= -\omega_{m} \left < q \right >+ G |\left <a \right >|^{2}- \gamma_{m} \left <p \right >,
	\end{equation}
	\begin{equation} 	
	\left < \dot{a} \right > =(-i\Delta_{c}-\kappa_{a}) \left <a \right >+ iG \left <a \right > \left < q \right >+ (E_{0}+\epsilon Cos(\Omega t))- i g_{0} \left < \sigma_{-} \right >,
	\end{equation} 
	\begin{equation}
	\left < \dot{\sigma_{-}} \right >=-(\kappa_{d}-i\frac{\Delta_{0}}{2}[1-cos(\omega_{e}t)] \left < \sigma_{z} \right >) \left < \sigma_{-} \right > + ig_{0} \left < a \right > \left < \sigma_{z} \right >.
	\end{equation}

	After applying the standard linearization technique to the set of Eqns.(2)-(5), we obtain the linearized quantum Langevin equations for the quantum fluctuations:

	\begin{equation}
	\dot{\delta q}=\omega_{m}\delta p,
	\end{equation}
	\begin{equation} 	
	\dot{\delta p}=-\omega_{m}\delta q+ G(\left < a \right > \delta a^{\dagger} + \left < a ^{\dagger} \right > \delta a)-\gamma_{m} \delta p+ \zeta(t),
	\end{equation}
	\begin{equation} 	
	\dot{\delta a}=(-i\Delta_{c}-\kappa_{a})\delta a+iG( \left < a \right > \delta q+ \left < q \right > \delta a) + \sqrt{2k_{a}}a_{in}- i g_{0}\delta\sigma_{-},
	\end{equation} 
	\begin{equation}
	\dot{\delta\sigma_{-}}=-(\kappa_{d}-i\frac{\Delta_{0}}{2}[1-cos(\omega_{e}t)] \left < \sigma_{z} \right >)\delta\sigma_{-} + ig_{0}\delta a \left < \sigma_{z} \right >,
	\end{equation}
	
	In the following, we will denote $\left < \sigma_{z}\right >$ as $N$. We now proceed ahead in the next section to understand the dynamics of the mean values and that of the fluctuations. To illustrate the analytical and numerical results, we chose realistic parameters which are accessible in experiments.
	
	\section{DYNAMICAL BEHAVIOR OF THE SYSTEM}
	
	The evolution of the system described above is obtained by evaluating the dynamics of the mean-field values described by Eqns. (6)-(9) and subsequently the dynamics of the quantum fluctuations described by Eqns. (10)-(13) which in turn depends on Eqns. (6)-(9). 
	
\subsection{Mean Field Dynamics}

	Eqns.(6)-(9) are nonlinear and can be solved numerically to obtain the time evolution of the mean values. If we are distant from optomechanical instabilities and multistabilities, the optomechanical coupling can be treated in a perturbative way. Since $E(t+\tau)=E(t)$ and $\Delta_{d}(t+\tau)=\Delta_{d}(t)$, the asymptotic solutions of the mean values will have the same periodicity as that of the implemented modulation i.e $2 \pi/\Omega$. Thus, this justifies performing a double expansion of the mean-field values, $ \left < O(t) \right >$ (O=q,p,a,$\sigma$) in powers of optomechanical coupling constant $G$ and Fourier series:

	\begin{figure}[ht]
		\hspace{-0.2cm}
		\begin{tabular}{cccc}
			\includegraphics[scale=0.45]{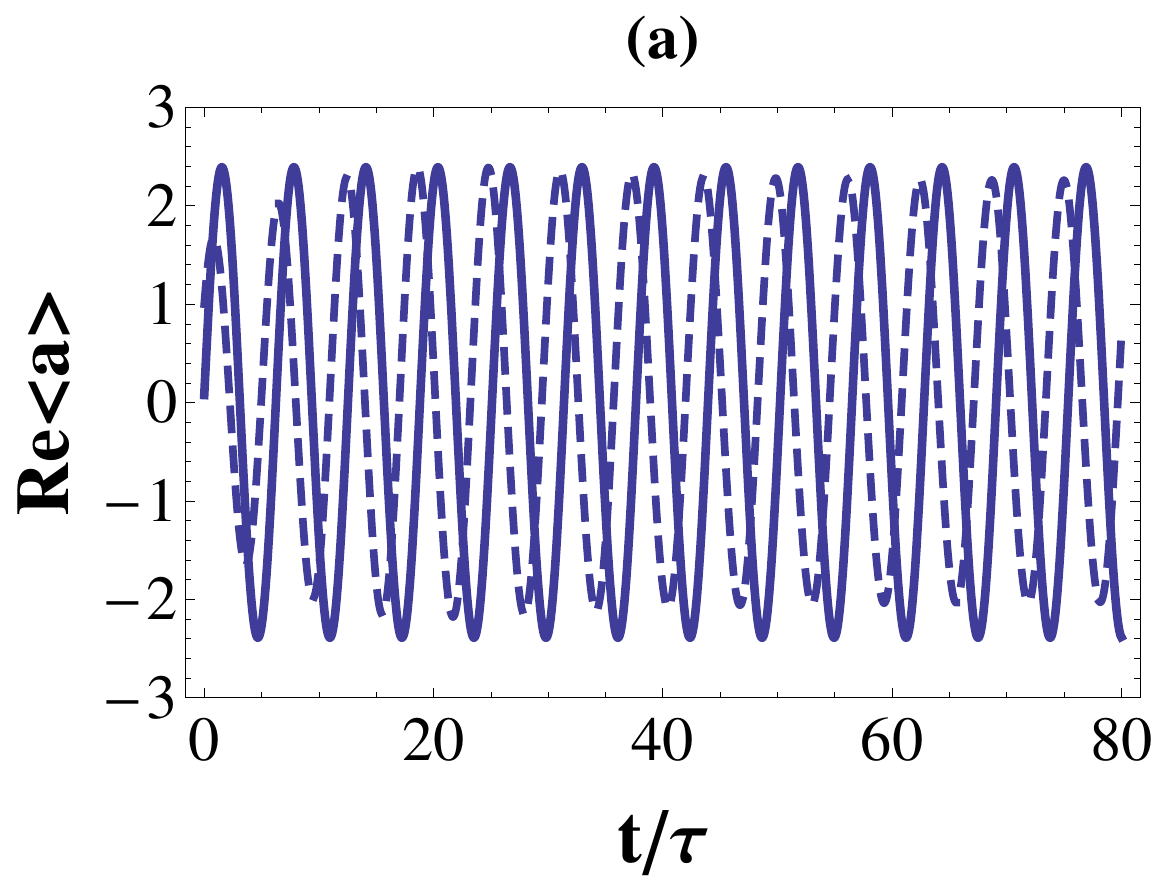}  \includegraphics[scale=0.45] {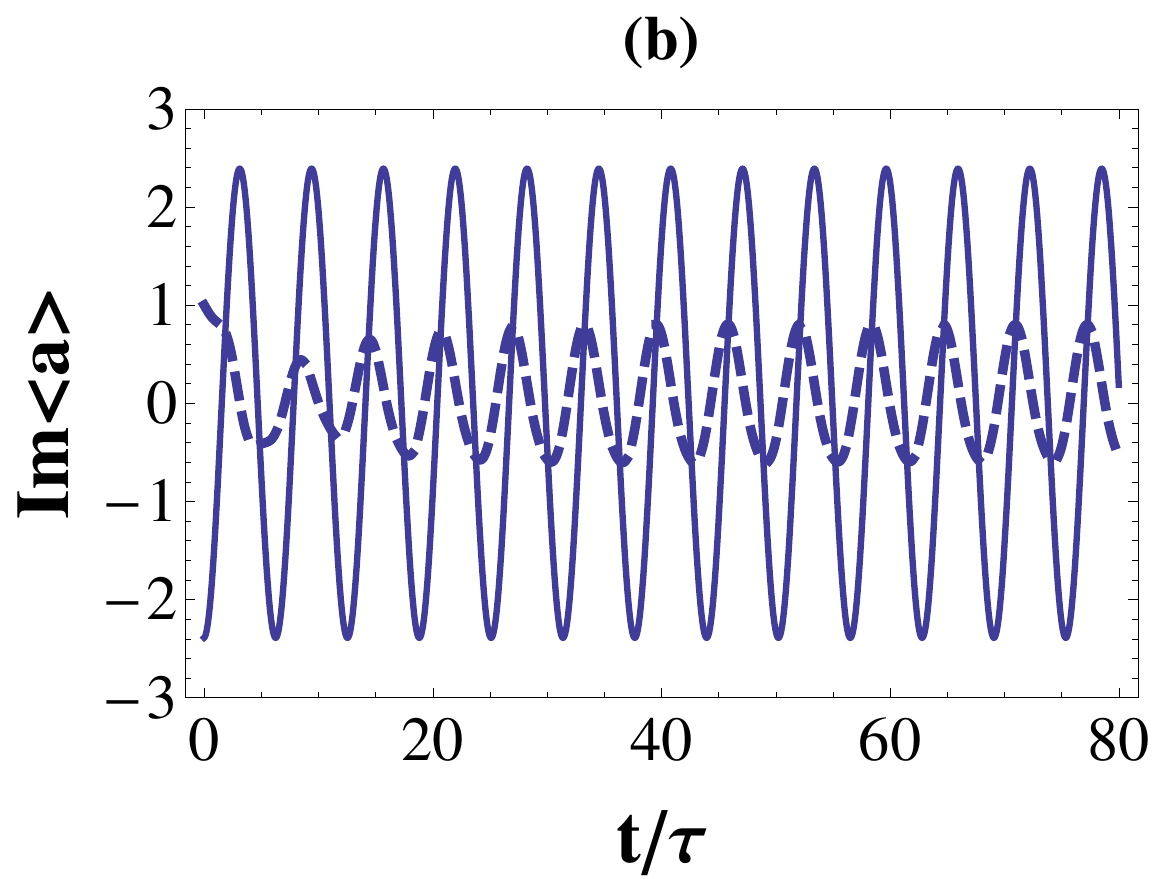}\\
			\includegraphics[scale=0.49]{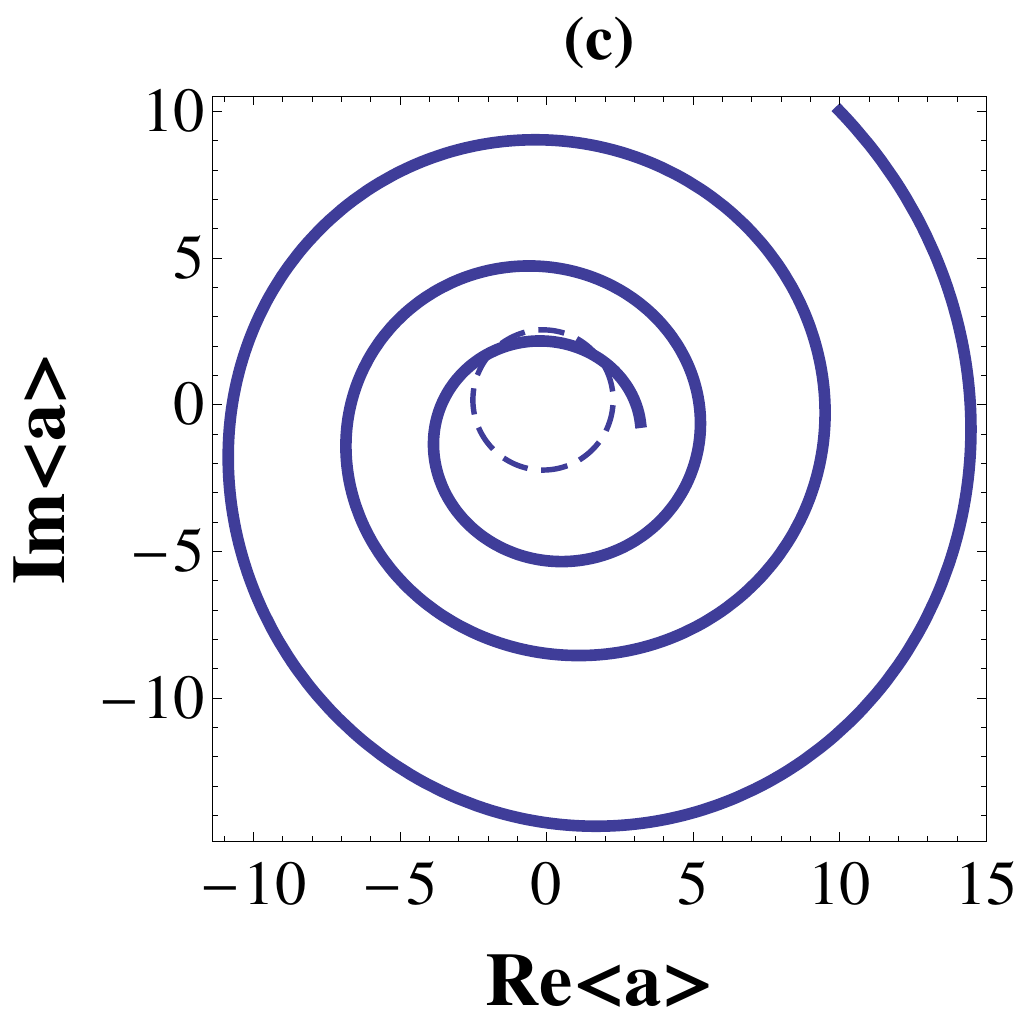}  \includegraphics[scale=0.45]	{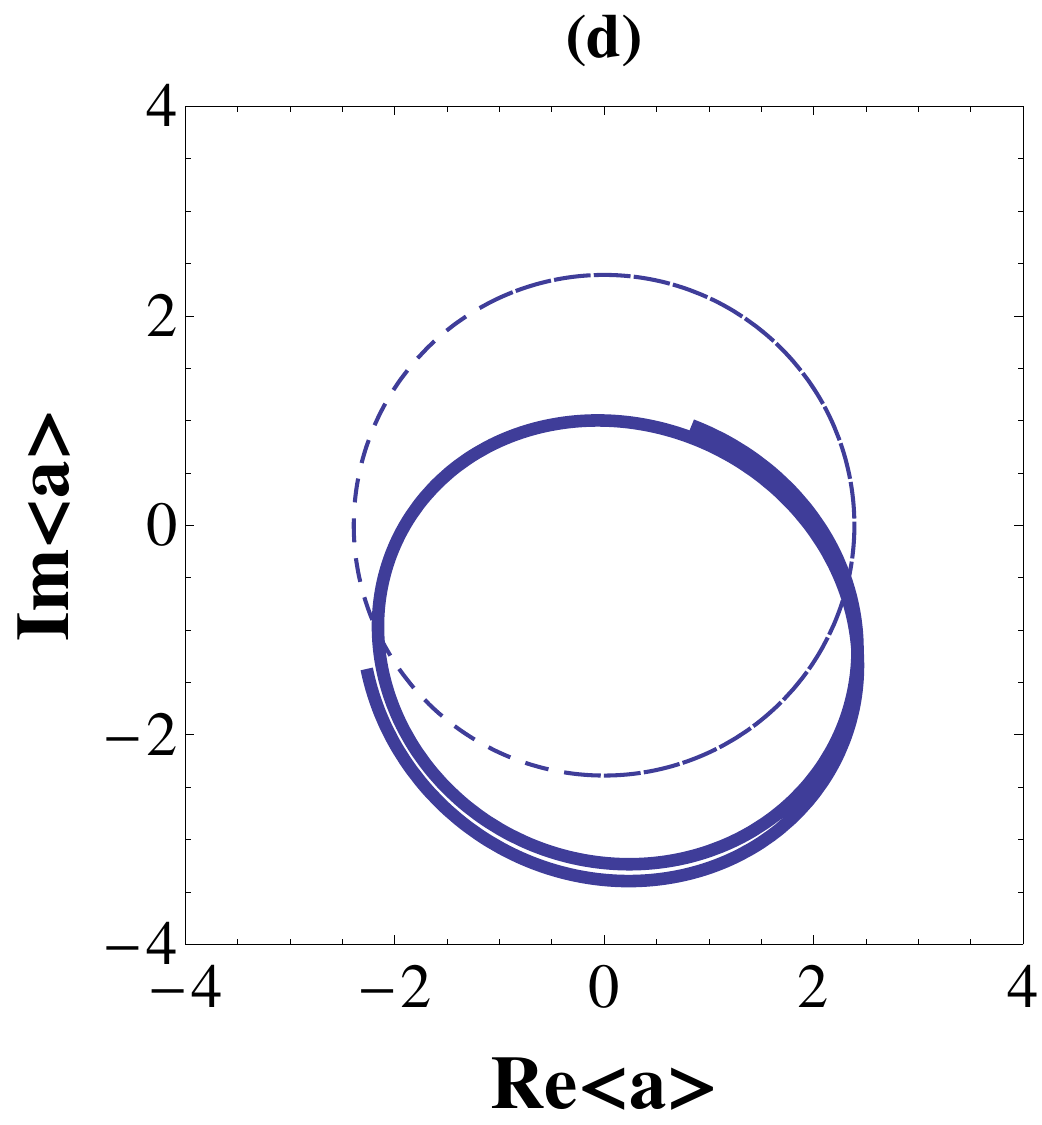}\\
			\includegraphics[scale=0.45]{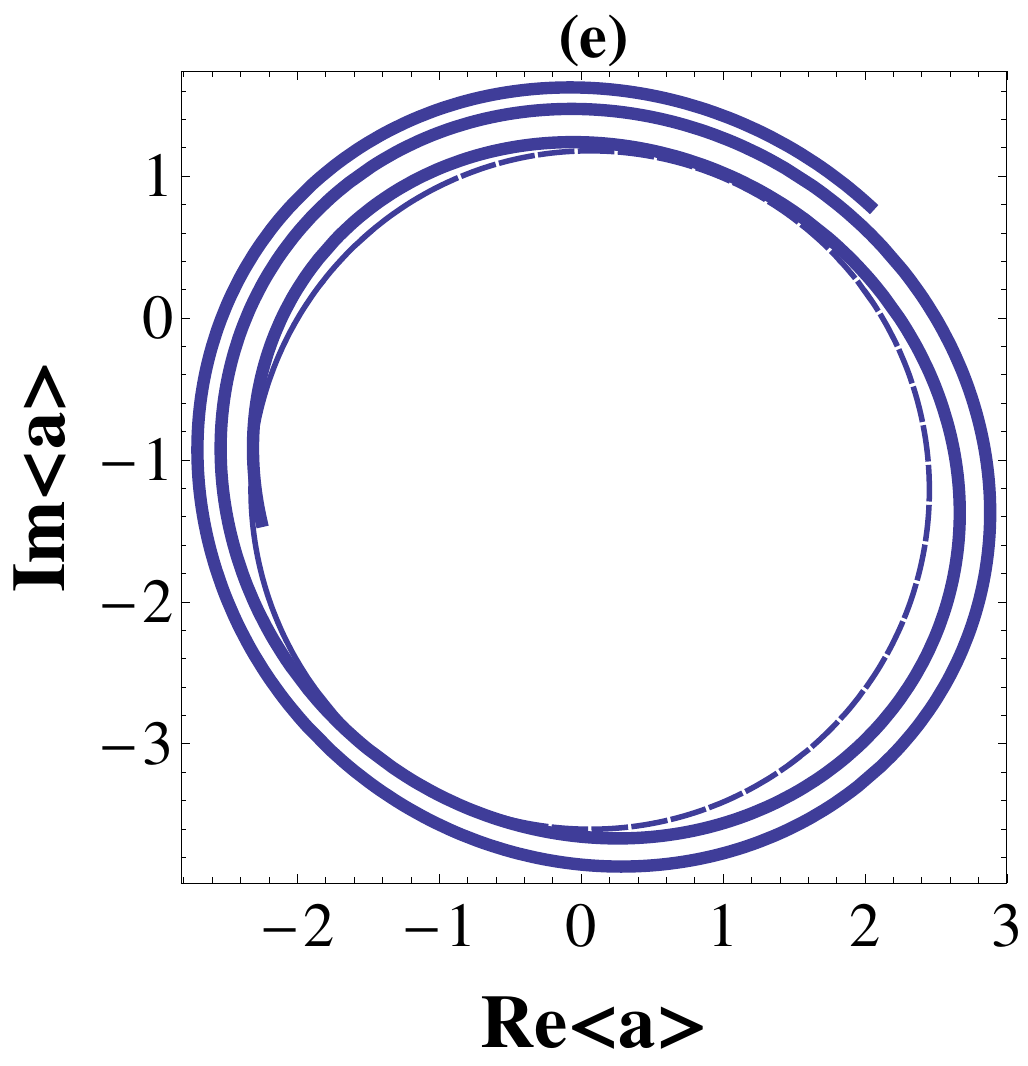}\\
		\end{tabular}
		\caption{(Color online): Time evolution of the real Re$ \left < a(t) \right >$ [plot (a)] and imaginary part Im$ \left < a(t) \right >$ [plot (b)]of the cavity mode mean value. Numerical (thick line) and analytical (thin-dotted line) results for phase space trajectories of cavity mode $ \left < a(t) \right >$ for the time intervals  (c): t =[0,25 $\tau$], (d): t=[25 $\tau$,35 $\tau$] and (e): t=[35 $\tau$,50 $\tau$]. The chosen parameters in units of $\omega_m$ are: $g_0$=0.3, N=1, $G$=0.01, $\gamma_{m}$=0.01, $E_0$=1, $E_1$= $E_{-1}$= 5, $\kappa_a$=0.1, $\epsilon$= 0.6, $\Delta_0$=1, $\kappa_d$=0.2, $\Omega$=1, and $\Delta_c$=1.}
	\end{figure}

	\begin{equation}
	<O(t)> = \sum_{j=0}^{\infty}\sum_{n=-\infty}^{\infty} O_{n.j}e^{in\Omega t}G^{j},
	\end{equation}

	where, $\Omega$=2 $\pi/\tau$ is the fundamental frequency of the modulation. We will assume that the external field and the QD detuning have the same modulation frequency $\Omega$.
	
	Now, substitute equation (14) in Eqn. (6-9) and also note similar Fourier series of the periodic driving amplitude $E(t)$ and QD detuning $\Delta_{d}(t)$,
	
	\begin{equation}
	E(t) = \sum_{n=-\infty}^{\infty}E_{n}e^{in\Omega t},
	\end{equation}

	\begin{equation}
	\Delta_{d}(t) = \sum_{n=-\infty}^{\infty}\Delta_{0}e^{in\Omega t},
	\end{equation}
	
	Thus one finds that the coefficients $O_{n,j}$(time independent) can be calculated by the following recursive relations:-
	
	\begin{equation}
	\begin{split}
	p_{n,0} & = 0, \; \;
	q_{n,0}= 0, \; \;  a_{n,0} = \frac{E_{-n}(in\Omega_{\kappa_{d}})}{[(\kappa_{a} + i(\Delta_{c}+n\Omega))(i n\Omega+\kappa_{d}) - g_{0}^2 N]}, \\
	\sigma_{n,0} &= \frac{E_{-n}(ig_{0}N)}{[(\kappa_{a}+i(\Delta_{c}+n\Omega))(in\Omega+\kappa_{d})-g_{0}^2 N]}, \\
	\end{split}
	\end{equation}
	
	associated with zeroth order perturbation with respect to $G$ and for all $j \geq 1$ we have
	
	\begin{equation}
	p_{n,j} = \frac{in\Omega q_{n,j}}{\omega_{m}}, \; \; q_{n,j} = \omega_{m}\sum_{k=0}^{j-1}\sum_{m=-\infty}^{\infty}\frac{a_{m,k}^{\dagger}a_{n+m,j-k-1}}{[\omega_{m}^2-(n\Omega)^2+i\gamma n\Omega]}
	\end{equation}
	
	\begin{equation}
	a_{n,j} = i\sum_{k=0}^{j-1}\sum_{m=-\infty}^{\infty}\frac{a_{m,k}q_{n-m,j-k-1}([\kappa_{a}+i(\Delta_{c}+n\Omega)][i(n\Omega-\Delta_{0})+\kappa_{a}])}{([\kappa_{a}+i(\Delta_{c}+n\Omega)][i(n\Omega-\Delta_{0})+\kappa_{d}]-g_{0}^2N},
	\end{equation}
	
	\begin{equation}
	\sigma_{n,j}=
	-g_{0}N\sum_{k=0}^{j-1}\sum_{m=-\infty}^{\infty}\frac{a_{m,k}q_{n-m,j-k-1}([\kappa_{a}+i(\Delta_{c}+n\Omega)]}{([\kappa_{a}+i(\Delta_{c}+n\Omega)][i(n\Omega-\Delta_{0})+\kappa_{d}]-g_{0}^2N)},
	\end{equation}
	
	corresponding to the $j^{th}$- order coefficients. In all the above calculations, we truncated the series to the terms with subscripts $|n| \leq3$ and $j \leq 4$. This truncation leads to a reasonable precision. All frequencies are dimensionless w.r.t $\omega_{m}$.
	
	\begin{figure}[pt]
		\hspace{-0.2cm}
		\begin{tabular}{cc}
			\includegraphics [scale=0.40]{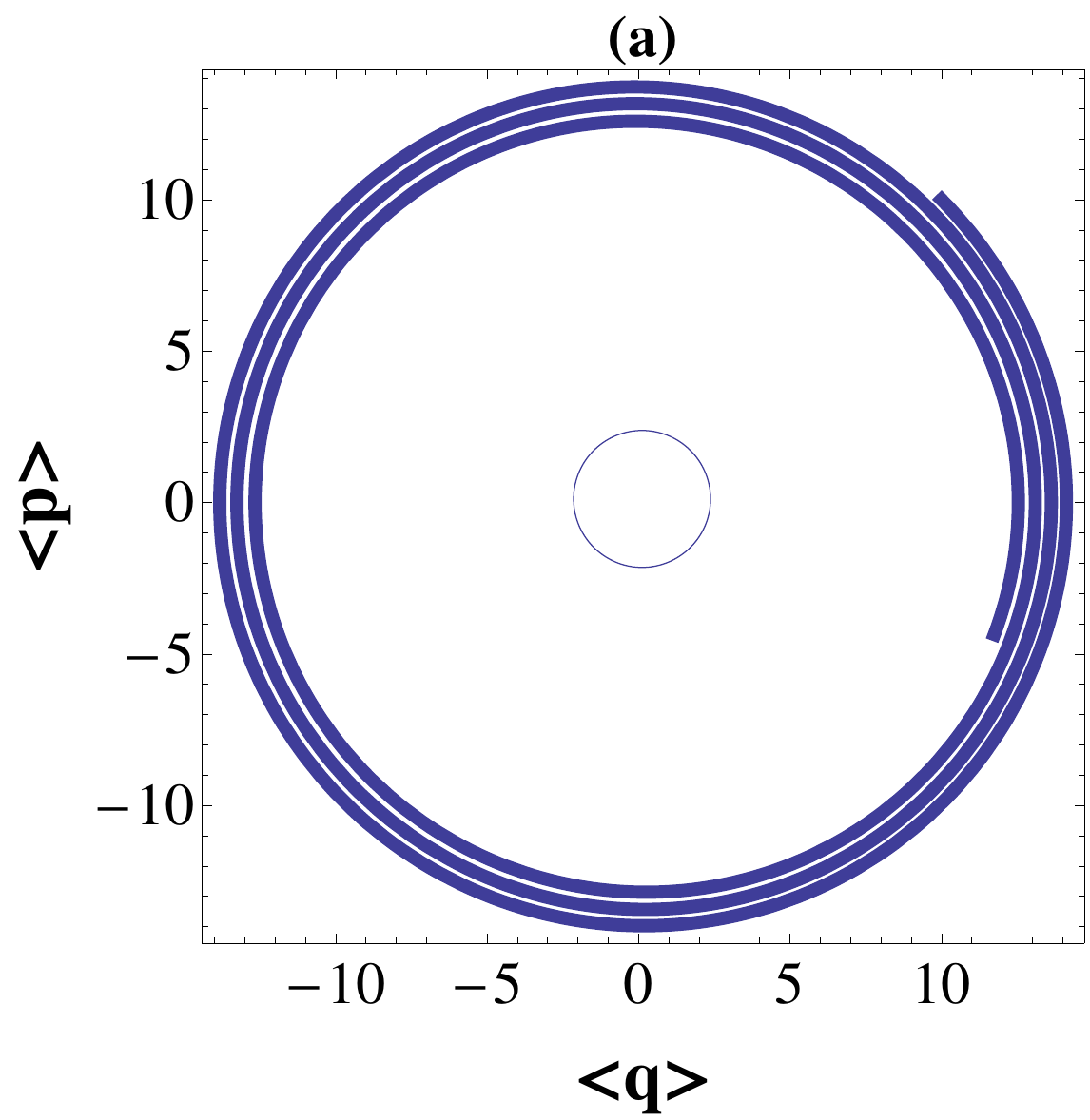}  \includegraphics [scale=0.40] {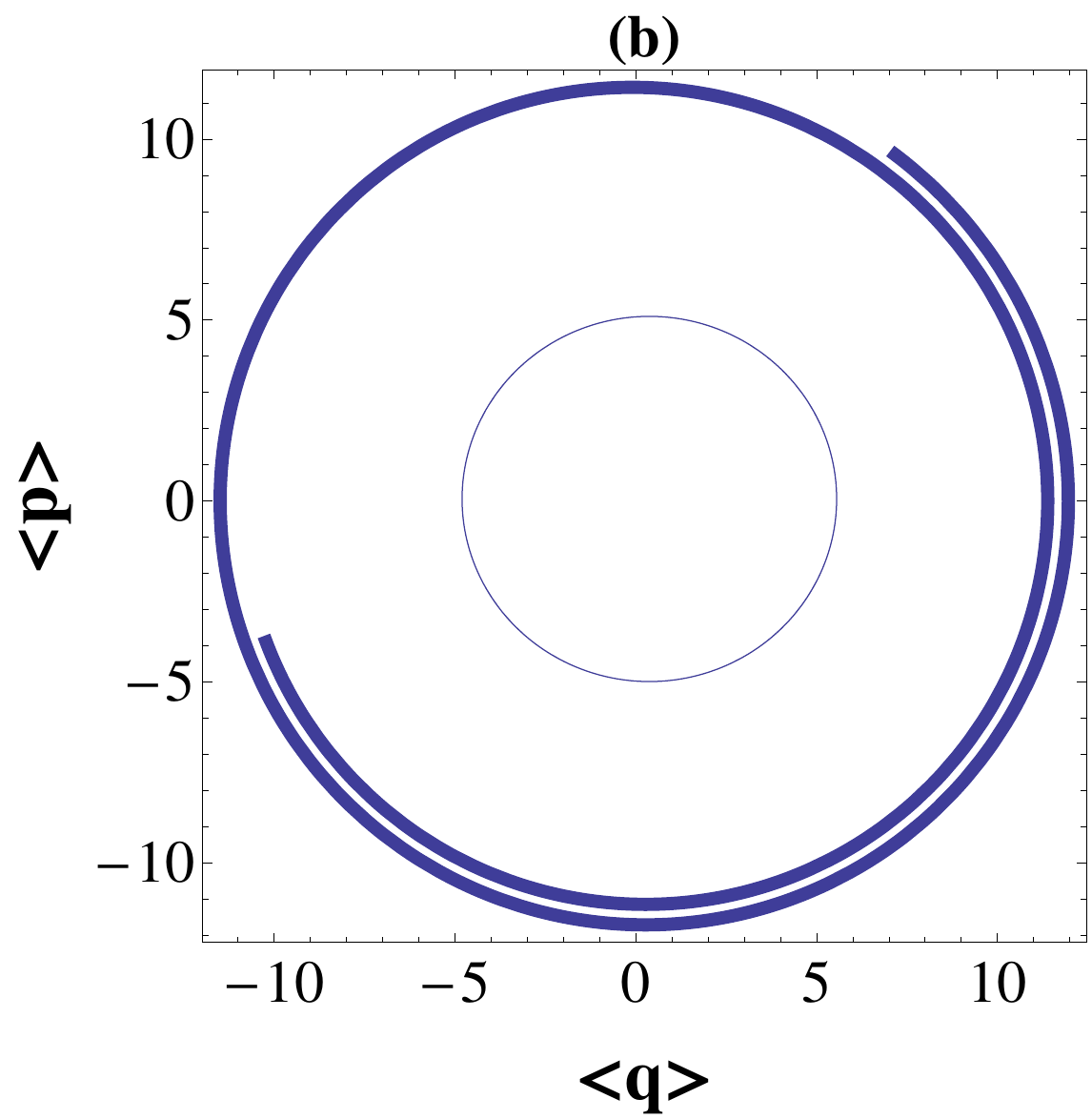}\\
			\includegraphics [scale=0.40]{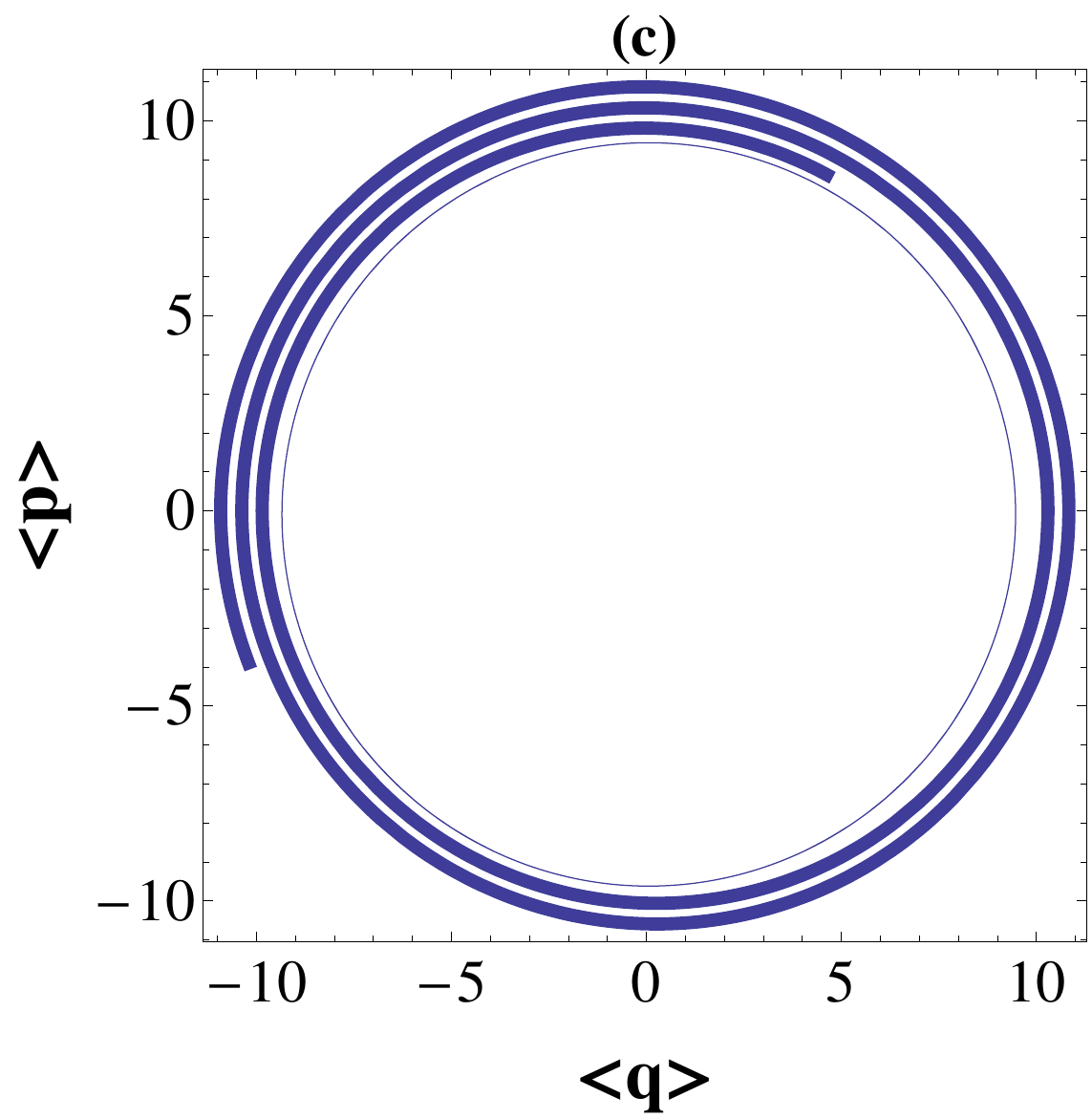} \\
		\end{tabular}
		\caption{(Color online) Phase space trajectories of dimensionless mechanical position and momentum mean values $<q(t)>$ and $<p(t)>$. Numerical simulation (thick line) and analytical results (thin-dotted line) for the time intervals: (a) t=[0,25$\tau$], (b): t=[25$\tau$, 35$\tau$] and (c): t=[35$\tau$,50$\tau$]. All the other parameters are same as those in Fig.2.}
		\label{fig:2}
	\end{figure}
	
	In Fig.2(a) and (b), we plot both the analytical and numerical results of the time evolution of the real ( Re$<a(t)>$ ) and imaginary ( Im$<a(t)>$ ) parts of the cavity mode mean values. We find that in the limit of long time, $ \left < a(t) \right >$ has the periodicity of $\tau(=2\pi/ \Omega)$ and that the numerical(thick line) and analytical results(thin-dotted  line) are in good agreement after about 40 modulation periods.To gain more insight about the dynamics, we plot the phase-space trajectory of cavity mode  $ \left < a(t) \right >$ from t=0 to t=50$\tau$. Figs.2(c), (d) and (e) shows phase space trajectories in the time interval, ($0 \rightarrow 25 \tau $),  ($25 \rightarrow 35 \tau $) and  ($35 \rightarrow 50 \tau $) respectively. As expected, after about 50 modulation periods, the trajectories finally converge to a limit cycle which is well approximated by the corresponding analytical analysis, as evident in Fig. 2(e). In Fig.3, we plot the phase space trajectories of the dimensionless mechanical position and momentum mean values $\left < q(t) \right>$ and $\left < p(t) \right >$. We again find that mean values $\left < q(t) \right>$ and $\left < p(t) \right >$ reach the analytically expected limit cycle after about 50 modulation periods. Thus one can design the modulation of the driving laser and the QD detuning to realize the mean values of the cavity mode, mechanical position and momentum with any periodic form. In this work, we would be interested in the mean values which acquires the same period of modulation in the long time limit.
	
	\subsection{Fluctuation Dynamics}
	
	Once we have obtained the dynamics of the mean values, now the dynamics of the quantum fluctuations can be solved. To proceed further,  we ignore the second and higher order small terms in Eqns.(10)-(13). We write the field operators in quadrature form as $\delta x=\frac{1}{\sqrt{2}}\left(\delta a^{\dagger} + \delta a \right)$,   $\delta y=\frac{1}{\sqrt{2}i}\left(\delta a^{\dagger} - \delta a \right)$, $\delta v=\frac{1}{\sqrt{2}}\left(\delta \sigma_{-}^{\dagger} + \delta \sigma_{-} \right)$, $\delta w=\frac{1}{\sqrt{2}i}\left(\delta \sigma_{-}^{\dagger} - \delta \sigma_{-} \right)$  and the quadratures of the noise operators as $\delta x^{in}=\frac{1}{\sqrt{2}}\left(\delta a^{in \dagger} + \delta a^{in}\right), \delta y^{in}=\frac{1}{\sqrt{2}i}\left(\delta a^{in\dagger}-\delta a^{in}\right)$,
	
	The quantum Langevin equations for these quadrature fluctuations can be written in a matrix form:

	\begin{equation}
	\dot{u}= D u + k,
	\end{equation}
	
	where, $u=(\delta q, \delta p, \delta x, \delta y,  \delta u, \delta v)^{T}$ is the fluctuation vector, $k = (0,\zeta,\kappa_{a},\kappa_{d},0,0)^{T}$ is the noise vector and the time dependent matrix is given by:
	
	\[
	D=
	\begin{bmatrix}
	0  &  \omega_{m} &  0  &  0  &  0  &  0 \\
	-\omega_{m} &  -\gamma_{m}& \sqrt{2}GRe(A) &  \sqrt{2}GIm(A)  &  0  & 0 \\
	-\sqrt{2}GIm(A) &  0 &  -\kappa_{a} & F_{1} &  0  &   g_{0} \\
	\sqrt{2}G_{1}Re(A) &  0 & -F_{1} &  -\kappa_{a} & -g_{0} & 0 \\
	0  &  0   &  0  & -g_{0}N &  -\kappa_{d}  &  -M N  \\
	0  &  0   &  g_{0}N  &  0  &  M N &   -\kappa_{d} \\ 
	\end{bmatrix},
	\]
	
	with

	\begin{equation}
	F_{1}=(\Delta_{c}-G Q),  \; \;
	M=\frac{\Delta_{0}}{2}\left(1-\cos{\omega_{e}t} \right).      
	\end{equation}
	
	Here, N=$<\sigma_{z}>$ is the difference of population in the ground state and excited state of Q.D \citep{wu,malinovsky}.
	
	\begin{figure}[ptb]
		\centering
		\begin{tabular}{@{}cccc@{}}
			\includegraphics[width=.40\textwidth]{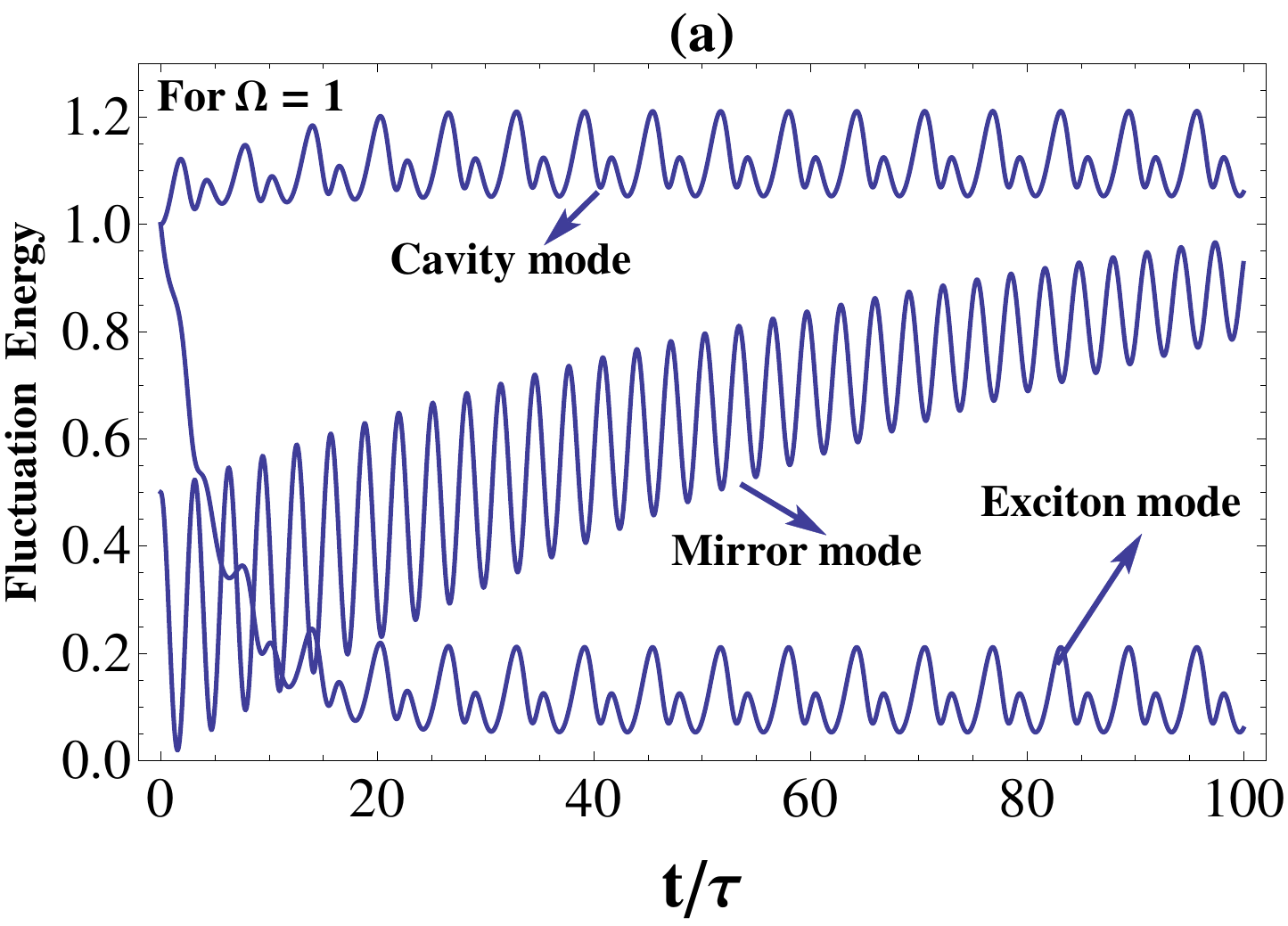}&
			\includegraphics[width=.40\textwidth]{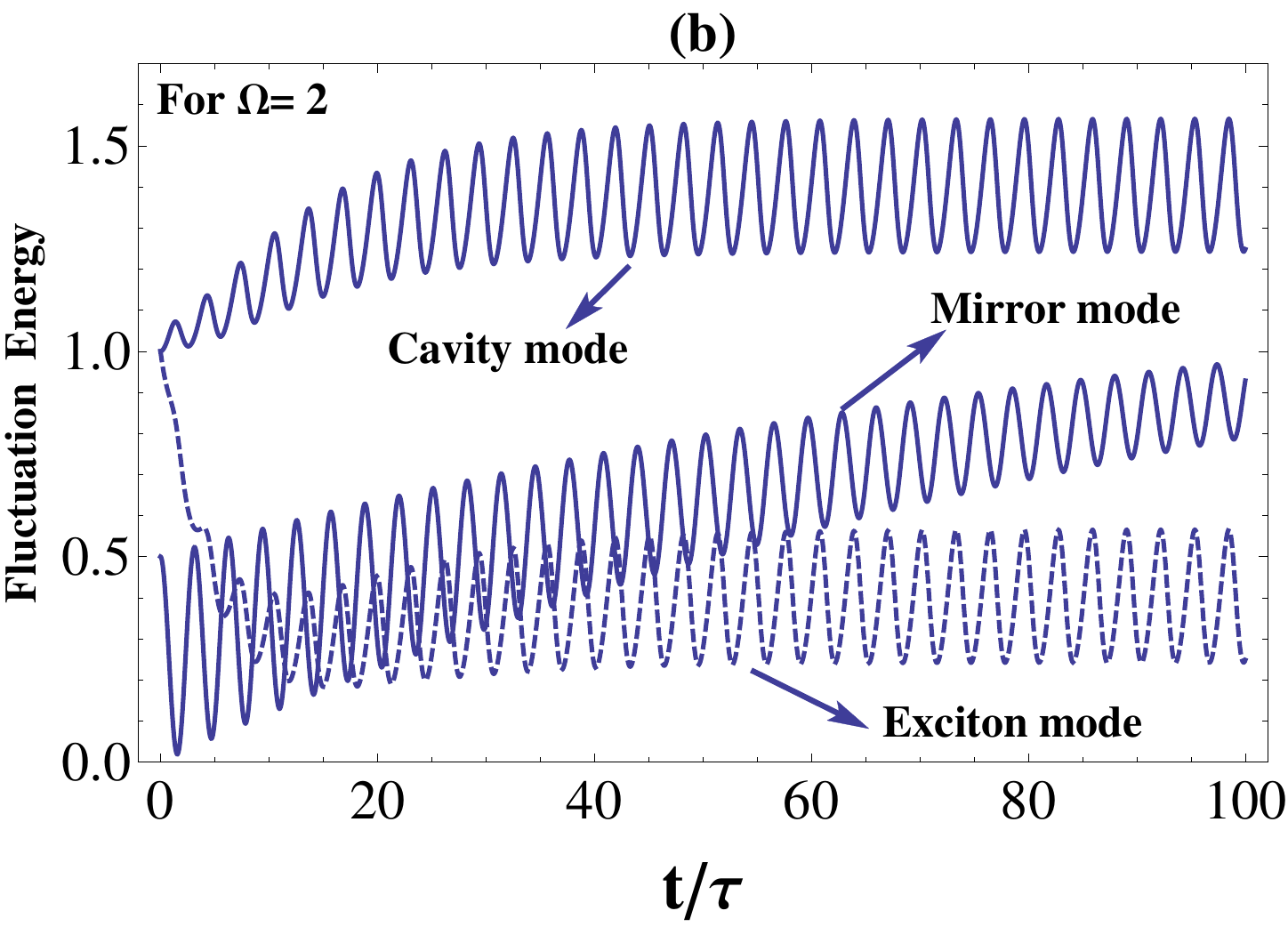} \\
			\includegraphics[width=.40\textwidth]{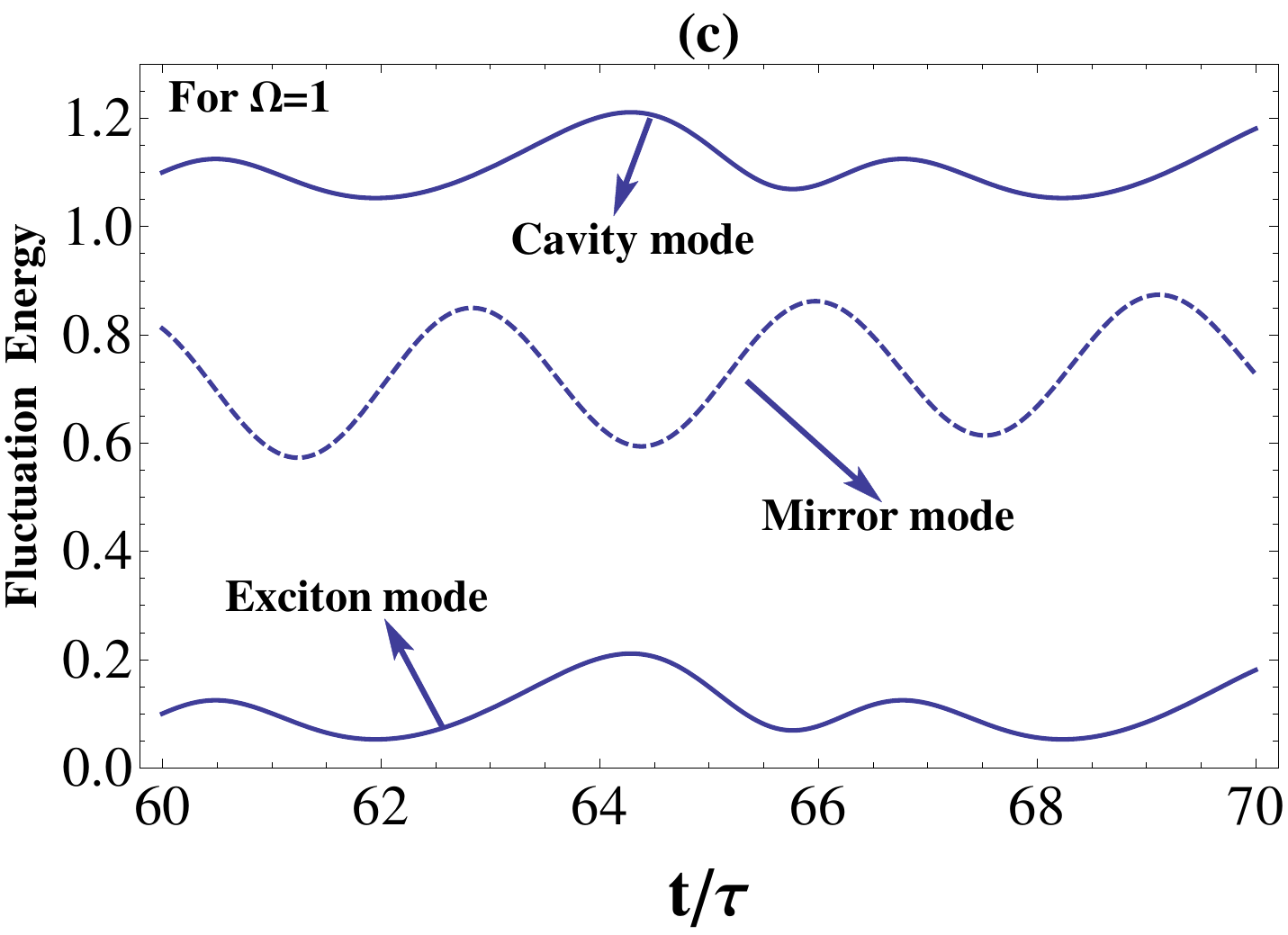}&
			\includegraphics[width=.40\textwidth]{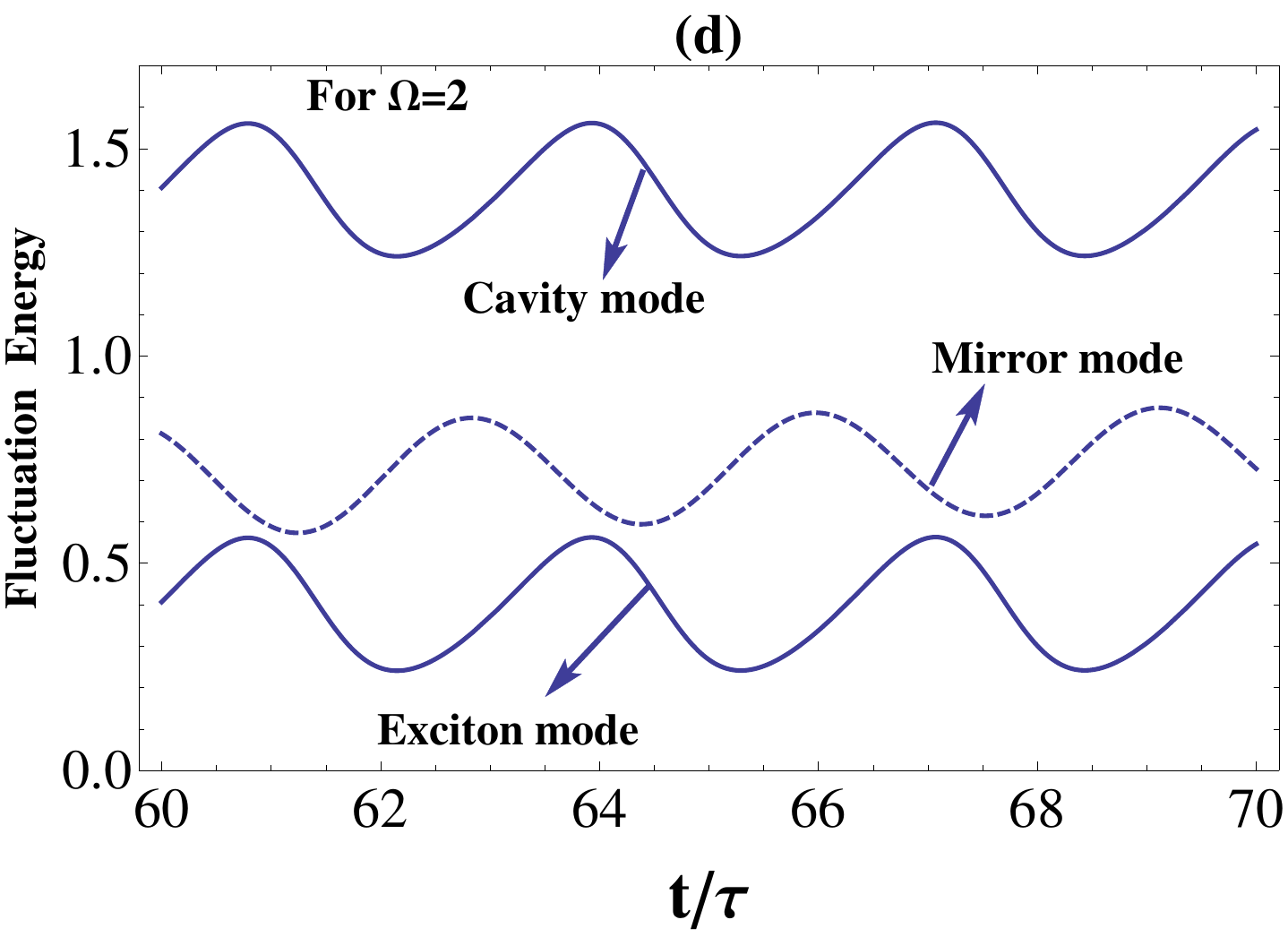} \\
			\end{tabular}
		\caption{(Color online) Time dynamics of energy fluctuations of various modes. The dynamics indicate energy transfer between different modes of the system. Plot (a) is for $\Omega=1$ while plot (b) is for $\Omega=2$. Plots (c) and (d) shows the long time behavior corresponding to $\Omega=1$ and $\Omega=2$ respectively. All other  parameters in units of $\omega_{m}$ are: $\kappa_{a}$=0.1, $E_{0}$=1, $\epsilon$=0.6, $\Delta_{0}$=1 $\kappa_{d}$=0.2, $\Delta_{c}$=1, $g_{0}$=0.3, $G$=0.005, N=1, $\gamma_{m}$=0.01.}
	\end{figure}
	
	Equation(21) is stochastic and needs some manipulation. Nonetheless, since we have linearized the dynamics and the noise terms are Gaussian zero-mean, fluctuations in the stable regime will also develop into asymptotic Gaussian zero-mean state. Instead, the correlation matrix V of elements completely defines the state of the system.
	
	\begin{equation}
	V_{ij}=\frac{1}{2}\left< u_{i}(t) u_{j}(t) + u_{j}(t)u_{i}(t)\right>,
	\end{equation}

	and attain its dynamic equation:-
	
	\begin{equation}
	\dot{V}= DV + VD^{T} + \tilde{N},
	\end{equation}
	
	where $D^{T}$ is transpose of $D$ and $\tilde{N}$ is the diagonal noise correlation matrix with diagonal elements (0,$\gamma_{m}(2n_{b}+1)$,$\kappa_{a}$,$\kappa_{a}$,0,$\gamma_{m}(2n_{b}+1)$,$\kappa_{d}$,$\kappa_{d}$). Now, equation(24) is an ordinary linear differential equation which gives thirty six equations in terms of $V_{ij}$. We can solve it numerically to study the dynamical behavior of the system. From Eqn. (24), fluctuation energy for the cavity mode is
  
\begin{equation} 
 \delta{a}^{\dagger}\delta{a}= \frac{(V_{33} + V_{44})}{2}
\end{equation}
 
Similarly for mirror mode, the fluctuation energy is given by $\delta q^{\dagger} \delta q = V_{11}$. For exciton mode we can derive the fluctuation energy as, 

\begin{equation}
\delta \sigma_{-}^{\dagger} \delta \sigma_{-} =\frac{(V_{55} + V_{66})}{2}.
\end{equation}

The value of $V_{ij}$ (where i,j =1,2,3,4,5) is evaluated by solving thirty six equations numerically. All relevant information about the system can be extracted directly from the correlation matrix V. In particular, we will focus on the following quantity: the average number of phonons in the mirror. The average number of phonons $\left < n_{phonon} \right >$, as we know depends on position and momentum of the moving mirror and this can be expressed using the approximate relation \citep{farace},
	
	\begin{equation}
	\hbar\omega_{m}\left(\left < n_{phonon} \right > + \frac{1}{2}\right) \approx \left(\frac{\hbar \omega_{m}}{2}\right)<\delta q^{2} + \delta p^{2}> = \left(\frac{\hbar \omega_{m}}{2}\right) (V_{11} + V_{22}),
	\end{equation}
	
	Eqn.(27) tells us how far the system is from the ground state.

	\begin{figure}[ht]
		\hspace{-0.2cm}
		\begin{tabular}{cc}
			\includegraphics[scale=0.40]{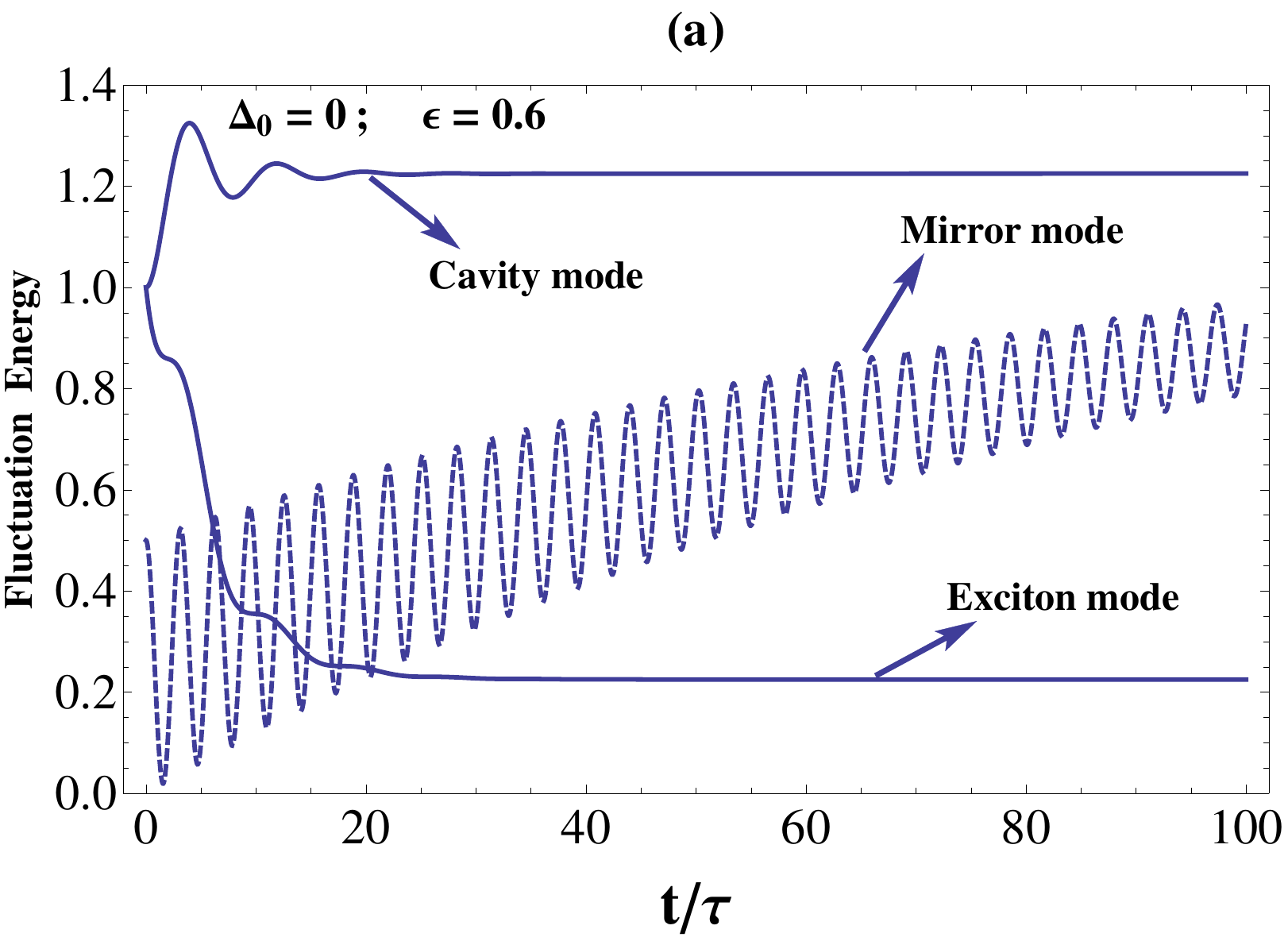}&\includegraphics[scale=0.40]{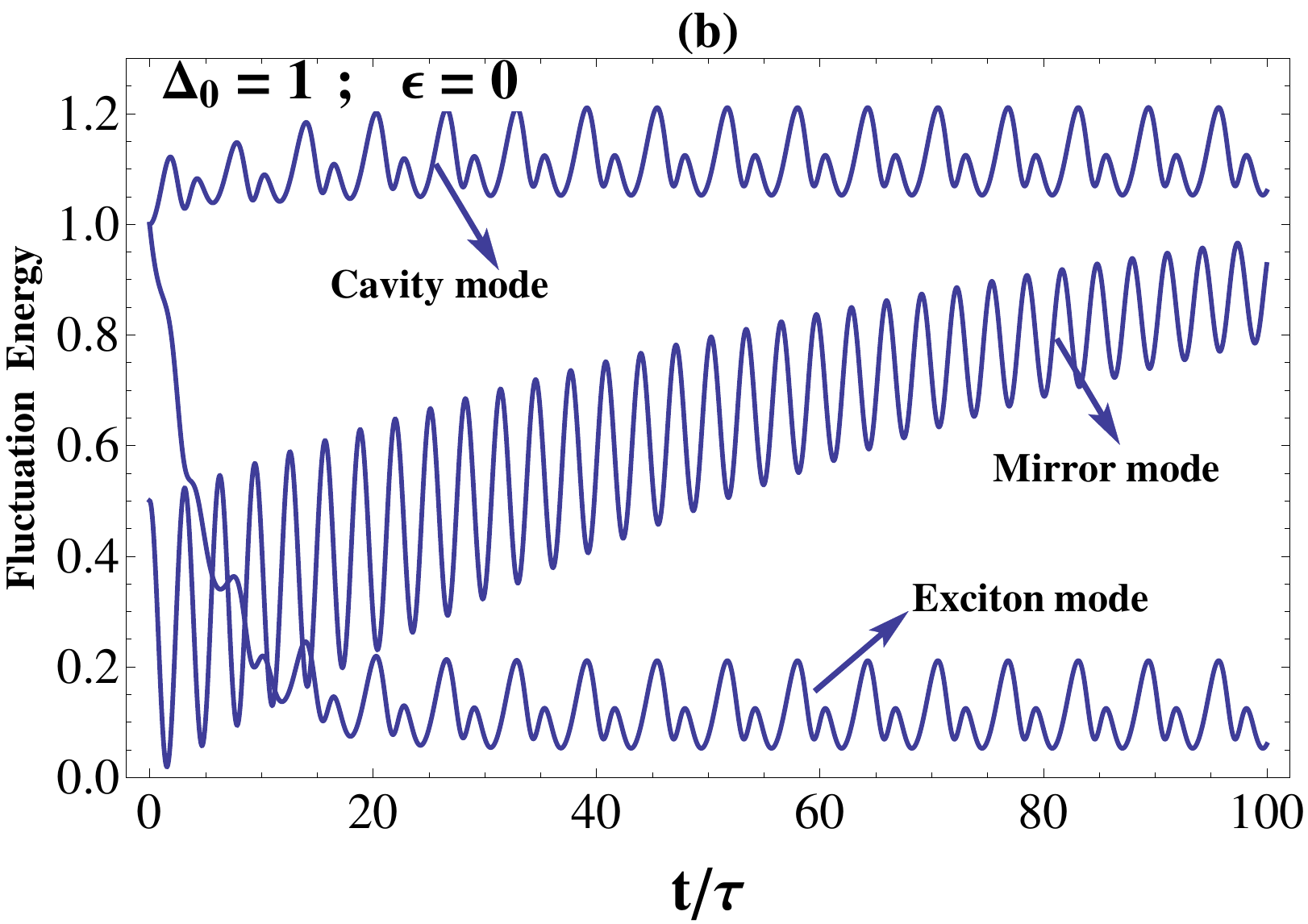} \\
		\end{tabular}
		\caption{(Color online) Temporal dynamics of the fluctuations when once the QD modulation is switched off (plot 5(a): $\Delta_{0}=0$, $\epsilon=0.6$) and once when the laser pump modulation is switched off (plot 5(b): $\Delta_{0}=1.0$, $\epsilon=0$).}
	\end{figure}

	\begin{figure}[ht]
		\hspace{-0.2cm}
		\begin{tabular}{c}
			\includegraphics[scale=0.60]{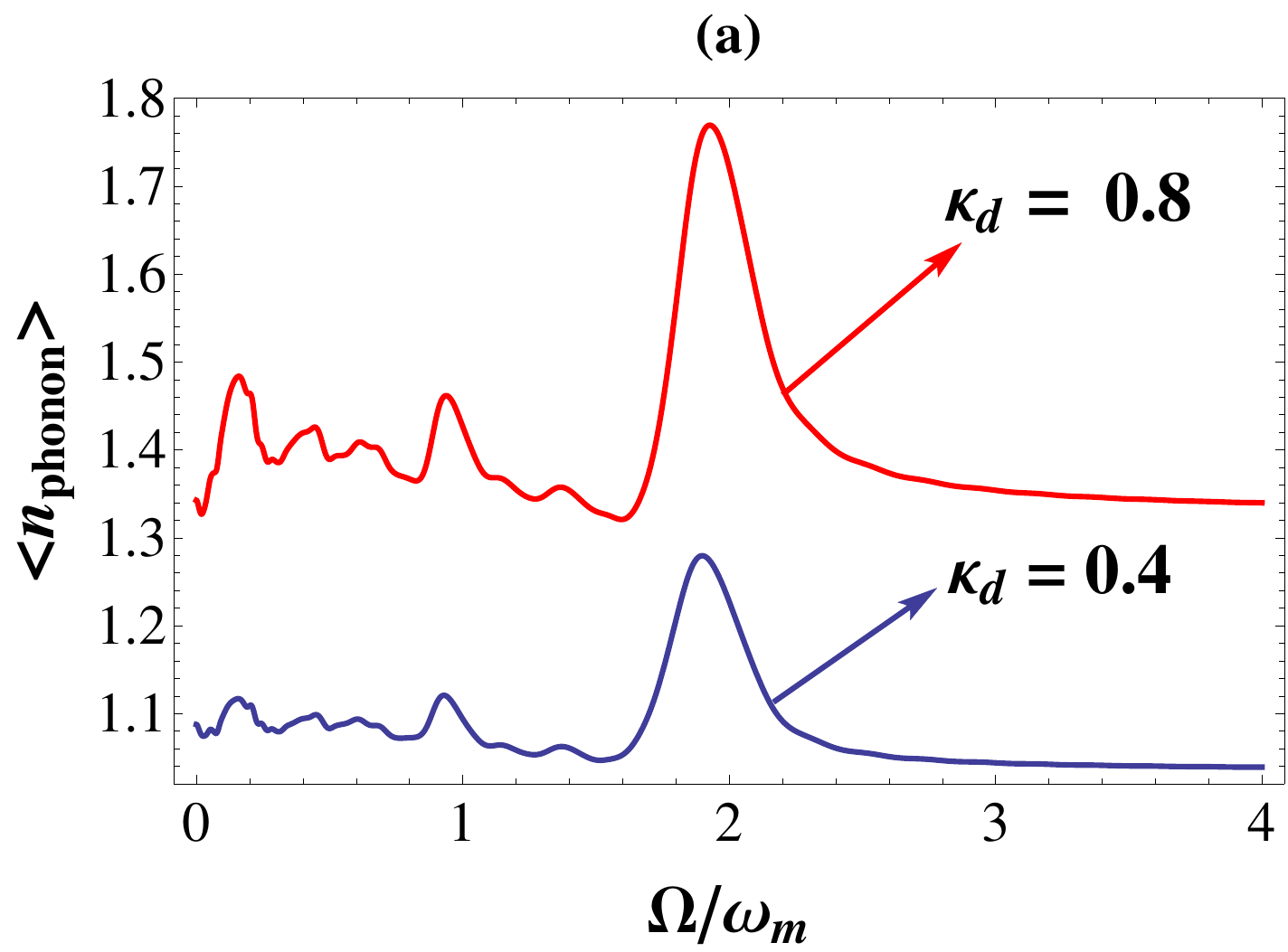} \\
		\end{tabular}
		\caption{(Color online)  Average number of phonons as a function of $\Omega$ for two different values of $\kappa_{d}$. The chosen parameters in units of $\omega_{m}$ are: $\kappa_{a}$=0.1, $E_{0}$=1, $\epsilon$=0.6, $\kappa_{d}$=0.2, $\Delta_{c}$=1, $g_{0}$=0.4,$\gamma_{m}$=0.00, N=1, $G$=0.1, $\Omega$=1. }
	\end{figure}

	To study the fluctuation energy transfer between different modes, we can employ a suitable laser pulse that would simultaneously couple the cavity mode to fluctuations of the exciton and the mirror. This would allow us to transfer the fluctuation energy from the mirror to the exciton via the cavity mode and vice versa, in a controlled and deterministic way for long times.  Unlike in stimulated Raman adiabatic passage (STIRAP) \citep{berg}, in our case, the intermediary cavity mode would be populated and so the pulse duration should be well within the decay timescale of the cavity mode.  In Fig. 4, we show how the fluctuation energies (dimensionless w.r.t $\omega_{m}$) can be transferred between exciton, mirror and cavity mode.
	We see in Fig. 4(a) and 4(b), that increasing the modulation frequency from $\Omega=1$ to $\Omega=2$ does not influence the mirror fluctuations. On the other hand, the fluctuation energy in the cavity and exciton modes is enhanced. The dynamics indicate that the energy that is pumped into the exciton and cavity mode is transferred to mirror mode. The exciton fluctuation energy can be transferred to the mirror mode indirectly through the cavity mode since the cavity mode is coupled to both the mirror as well as the exciton mode. There is a rapid drop in the exciton energy with a corresponding gradual rise in the cavity and mirror fluctuation energy which stabilizes for long times $t \approx 40 \tau$. This time interestingly is the same time during with the phase phase trajectories of the cavity and mirror mode converge to a limit cycle as found out in the previous section. In Fig.4(c) and (d), we show the plots of Fig.4(a) and (b) in the time duration $t=60 \tau \rightarrow 70 \tau$ respectively. We notice that the oscillations in the fluctuation energy of the cavity and exciton modes are in phase while the oscillations in the mirror mode is $\pi/2$ with respect to the other two modes. This essentially signifies that both the cavity and exciton mode can simultaneously pump (or extract) energy into(from) the mechanical mode. In Fig.5, we demonstrate the temporal dynamics of the fluctuations when once the QD modulation is switched off (plot 5(a): $\Delta_{0}=0$, $\epsilon=0.6$) and once when the laser pump modulation is switched off (plot 5(b): $\Delta_{0}=1.0$, $\epsilon=0$). We clearly note that the origin of the oscillations in the fluctuation energy after it reaches a saturation around $t \approx 20 \tau$ is due to the QD modulation. The mechanical quality factor also plays a significant role in the dynamics. It is observed (plots not shown) that the mechanical fluctuations of a high-Q oscillator reaches a steady value at a much later time compared to the case of a low-Q mechanical oscillator. This indicates that depending on the mechanical Quality-factor of the oscillator, a proper choice of the modulation is required to reach the stable fluctuation regime at the desired time. In Fig. 6, as the decay rate of the QD ($\kappa_{d}$) increases, the average number of phonons also increases which indicates that a rapidly decaying QD transfers higher energy to the mechanical mode.
  
  In this section we have shown how the fluctuations of various subsystems, namely, the cavity mode, the exciton mode and the mechanical mode coherently exchanges energy which can be controlled by the two modulations as well as the various other system parameters. This naturally leads us to the exploration of the entanglement between the fluctuations of the relevant continuous variables describing the various modes.

	\section{Entanglement Dynamics}
	Entanglement, a basic phenomenon of quantum mechanics  \citep{horo}, has been considered to be a primary tool for quantum computing, information processing and quantum communication \citep{kimble,rabl,braun,manc,xureb}. Many physical system have been prepared and manipulated so far, to study quantum entanglement such as individual atoms \citep{ritter}, photons \citep{kwiat,bowen}, ions \citep{jost} and solid-state spin qubit \citep{toga}. Entanglement has also been observed in an optomechanical system consisting of a quantum well embedded inside \citep{sete}. To study the entanglement there are different techniques, but Logarithmic negativity is one of the commonly used methods \citep{boura}.
	
	Equation (24) can be numerically solved with initial condition, V(0)= Diag[$n_{th}$+1/2,$n_{th}$+1/2,1/2,1/2,1/2,1/2]. In our study, we have assumed that a mechanical oscillator is prepared in its thermal state at the temperature T and the cavity field in its vacuum state. For entanglement between quantum dot and mirror, we make $4 \times 4$ matrix from the element $V_{ij}$ as,

	\begin{figure}[ht]
		\centering
		\begin{tabular}{@{}cccc@{}}
			\includegraphics[width=.40\textwidth]{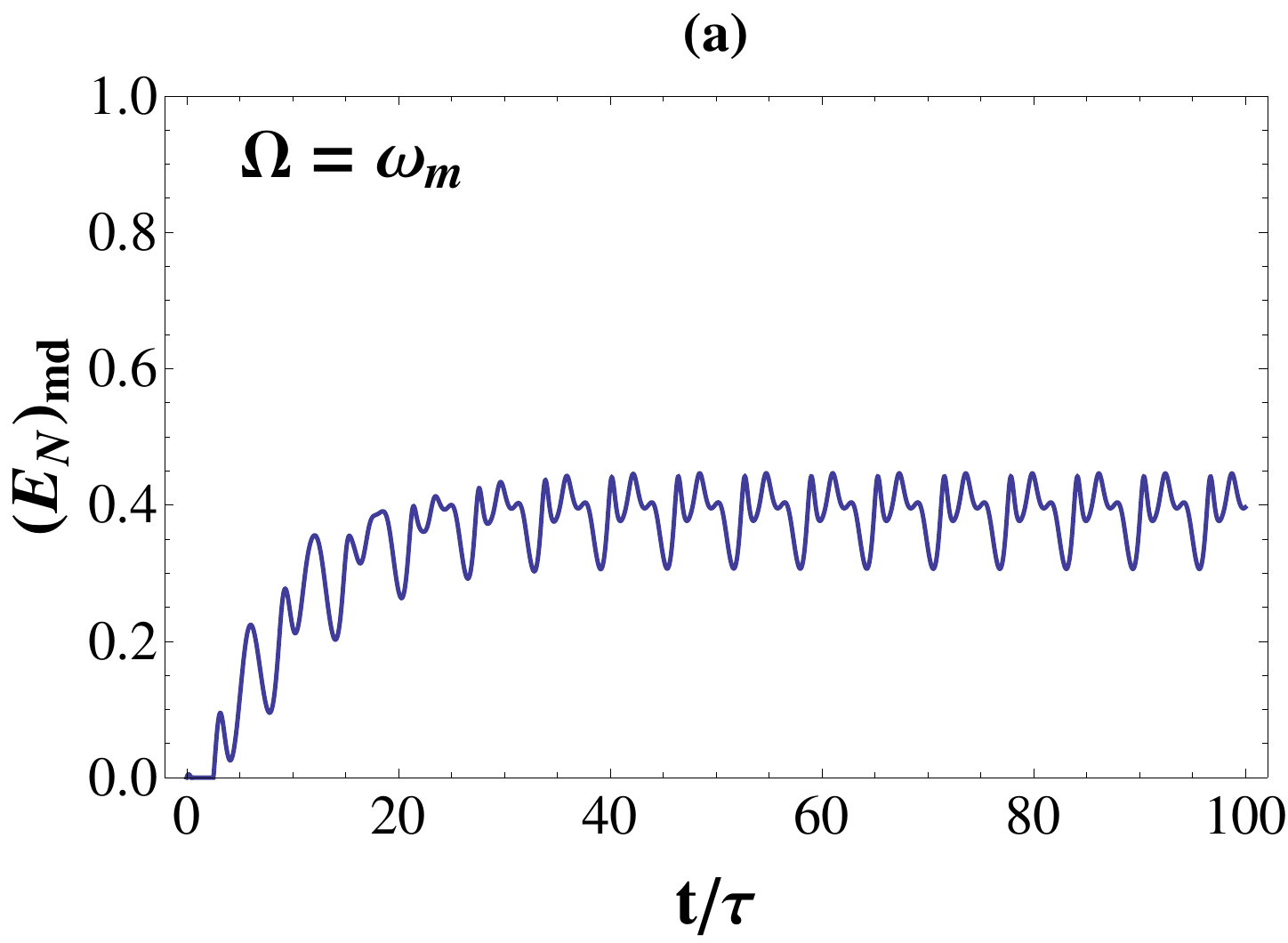}&
			\includegraphics[width=.40\textwidth]{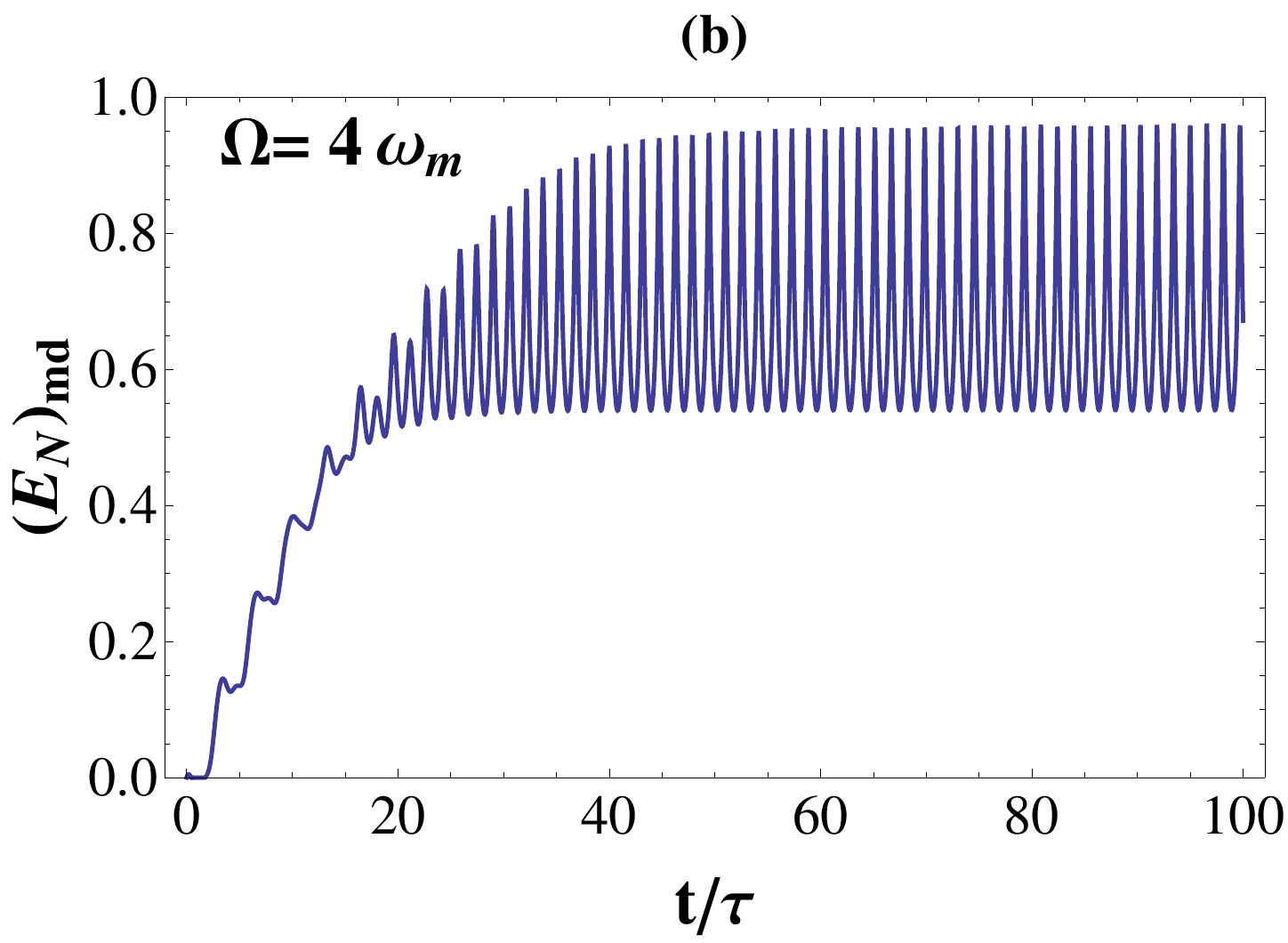} \\
		\end{tabular}
		\caption{(Color online) Dynamics of the QD- mirror entanglement $(E_{N})_{md}$ as a function of time for different modulation regime: (a). $\Omega$= $\omega_{m}$, (b). $\Omega$= $4\omega_{m}$;  The chosen parameters in units of $\omega_{m}$ are: $\kappa_{a}$=0.1, $E_{0}$=1, $\epsilon$=0.6, $\kappa_{d}$=0.2, $\Delta_{c}$=1, $g_{0}$=0.3, $n_{b}$=2, $\gamma_{m}$=0.01, N=1, $G$=0.005. }
	\end{figure}

 \[
 V^{`}=
 \begin{bmatrix}
 V_{11}  & V_{12} &  V_{15}  &  V_{16} \\
 V_{21}  & V_{22} &  V_{25}  &  V_{26} \\
 V_{31}  & V_{32} &  V_{55}  &  V_{56} \\
 V_{41}  & V_{42} &  V_{65}  &  V_{66} \\ 
 \end{bmatrix}.
 \]
 
The above matrix then can be rewritten in the following form:
 
 \[
 \begin{bmatrix}
 X & Z \\
 Z^{T} & Y \\
 \end{bmatrix},
 \]
 
 where X,Y and Z are 2 $\times$ 2 block matrices. The continuous variable entanglement between  two modes can be calculated by using the logarithmic negativity $E_{N}$ defined as,
 
 \begin{equation}
 E_{N}= max[0,-ln 2\nu^{-}],
 \end{equation}
 
 where $\nu^{-}$ = $\left(\Sigma(V^{`})-\sqrt{\Sigma(V^{`})^{2}-4 det V^{`}}\right)^{1/2}2^{-1/2}$ is the smallest symplectic eigenvalue of the partial transpose of $V^{`}$ with $\Sigma(V^{`})$= det(X) + det(Y) - 2det(Z). 

\begin{equation}
\text{det}(X) = [(V_{11}*V_{22})-(V_{21}*V_{12})]
\end{equation}

\begin{equation}
\text{det}(Y) = [(V_{55}*V_{66})-(V_{56}*V_{65})]
\end{equation}

\begin{equation}
\text{det}(Z) = [(V_{15}*V_{26})-(V_{16}*V_{25})]
\end{equation}

and

\begin{eqnarray}
\text{det}(V^{`}) &=& V_{11}V_{22}V_{55}V_{66} - V_{11}V_{22}V_{56}V_{65} - V_{11}V_{25}V_{52}V_{66} + V_{11}V_{25}V_{56}V_{62} \\ \nonumber
&+& V_{11}V_{26}V_{52}V_{65} - V_{11}V_{26}V_{55}V_{62} - V_{12}V_{21}V_{55}V_{66} + V_{12}V_{21}V_{56}V_{65} \\ \nonumber
&+&V_{12}V_{25}V_{51}V_{66}-V_{12}V_{25}V_{56}V_{61}-V_{12}V_{26}V{51}V_{65} + V_{12}V_{26}V_{55}V_{61} \\ \nonumber
 &+& V_{15}V_{21}V_{52}V_{66} - V_{15}V_{21}V_{56}V_{62} - V_{15}V_{22}V_{51}V_{66} + V_{15}V_{22}V_{56}V_{61} \\ \nonumber
  &+&V_{15}V_{26}V_{51}V_{62} - V_{15}V_{26}V_{52}V_{61} - V_{16}V_{21}V_{52}V_{65} + V_{16}V_{21}V_{55}V_{62} \\ \nonumber
  &+& V_{16}V_{22}V_{51}V_{65} - V_{16}V_{22}V_{55}V_{61} - V_{16}V_{25}V_{51}V_{62} + V_{16}V_{25}V_{52}V_{61}
\end{eqnarray}

We now proceed with the numerical investigation of  entanglement between the various modes of the system  (see appendix A).

 \begin{figure}[hp]
 	\hspace{-0.2cm}
 	\begin{tabular}{cc}
 		\includegraphics [scale=0.55]{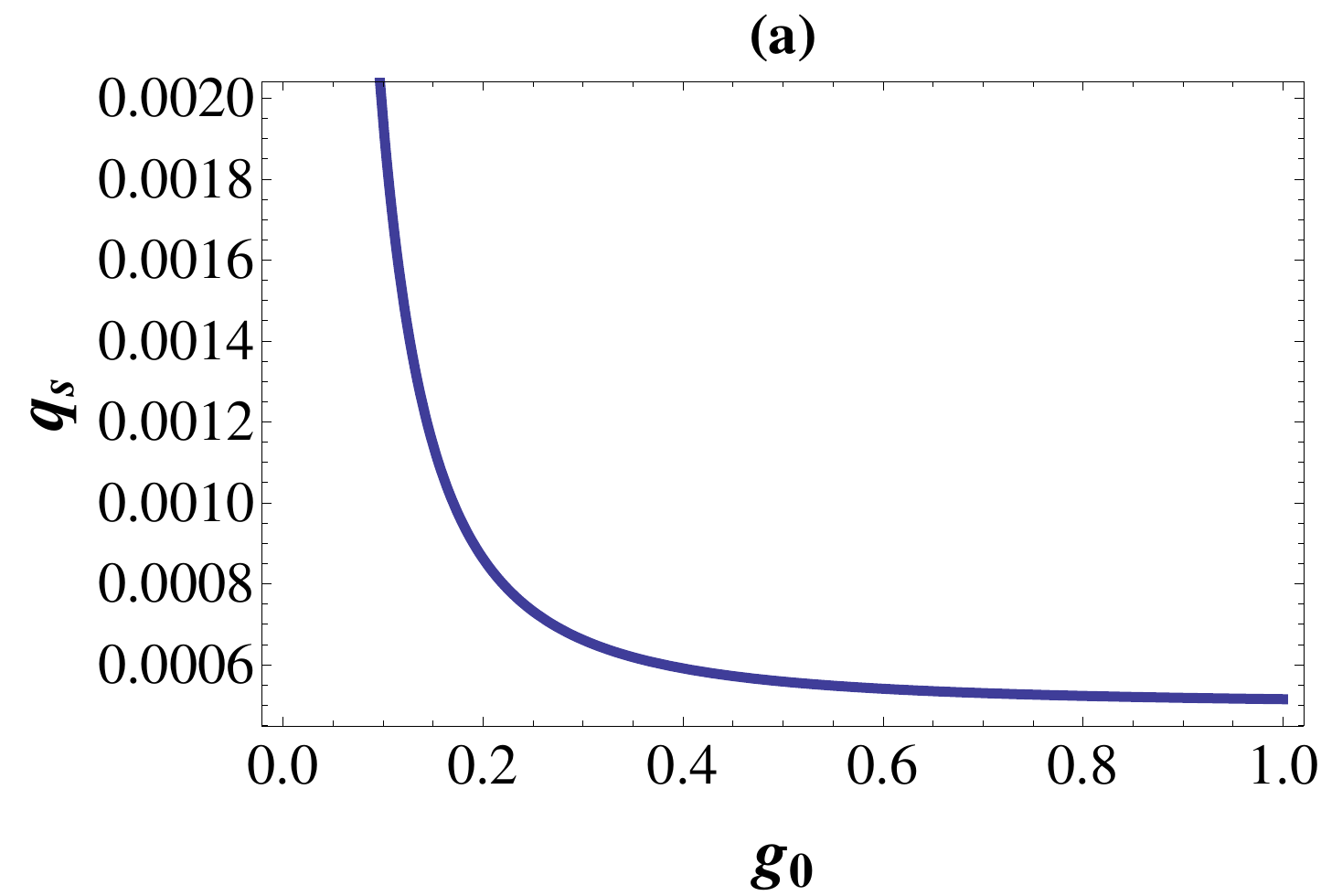}  \includegraphics [scale=0.37] {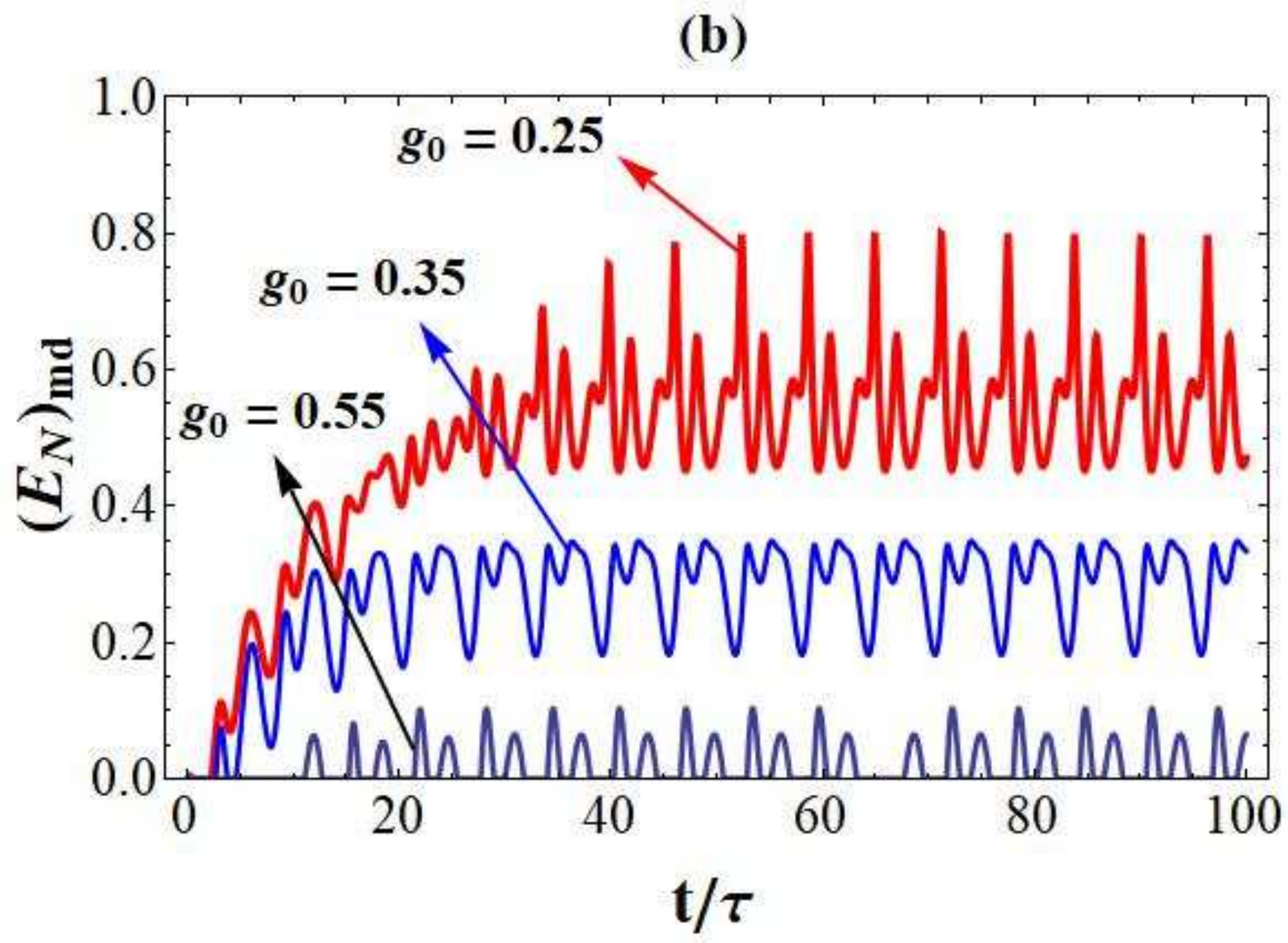}\\
 	\end{tabular}
  	\caption{(Color online)(a)- Steady state value of mirror position $q_{s}$ as a function of $g_{0}$. (b)- Dynamics of the QD- mirror entanglement $(E_{N})_{md}$ as a function of time for different values of $g_{0}$ $=$ 0.25(Red Solid), 0.35(Blue Dashed), 0.55(Green Thin Solid). The chosen parameters in units of $\omega_{m}$ are: $\Delta_{0}$=1, $G$=0.005, $\kappa_{d}$=0.2, $\kappa_{a}$=0.1, $\Delta_{c}$=1, $\epsilon$=0.6, $n_{b}$=2, $\gamma_{m}$=0.01, $E_{0}$=1, $\Omega$=1.}
\end{figure}

\begin{figure}[ht]
		\centering
		\begin{tabular}{@{}cccc@{}}
			\includegraphics[width=.40\textwidth]{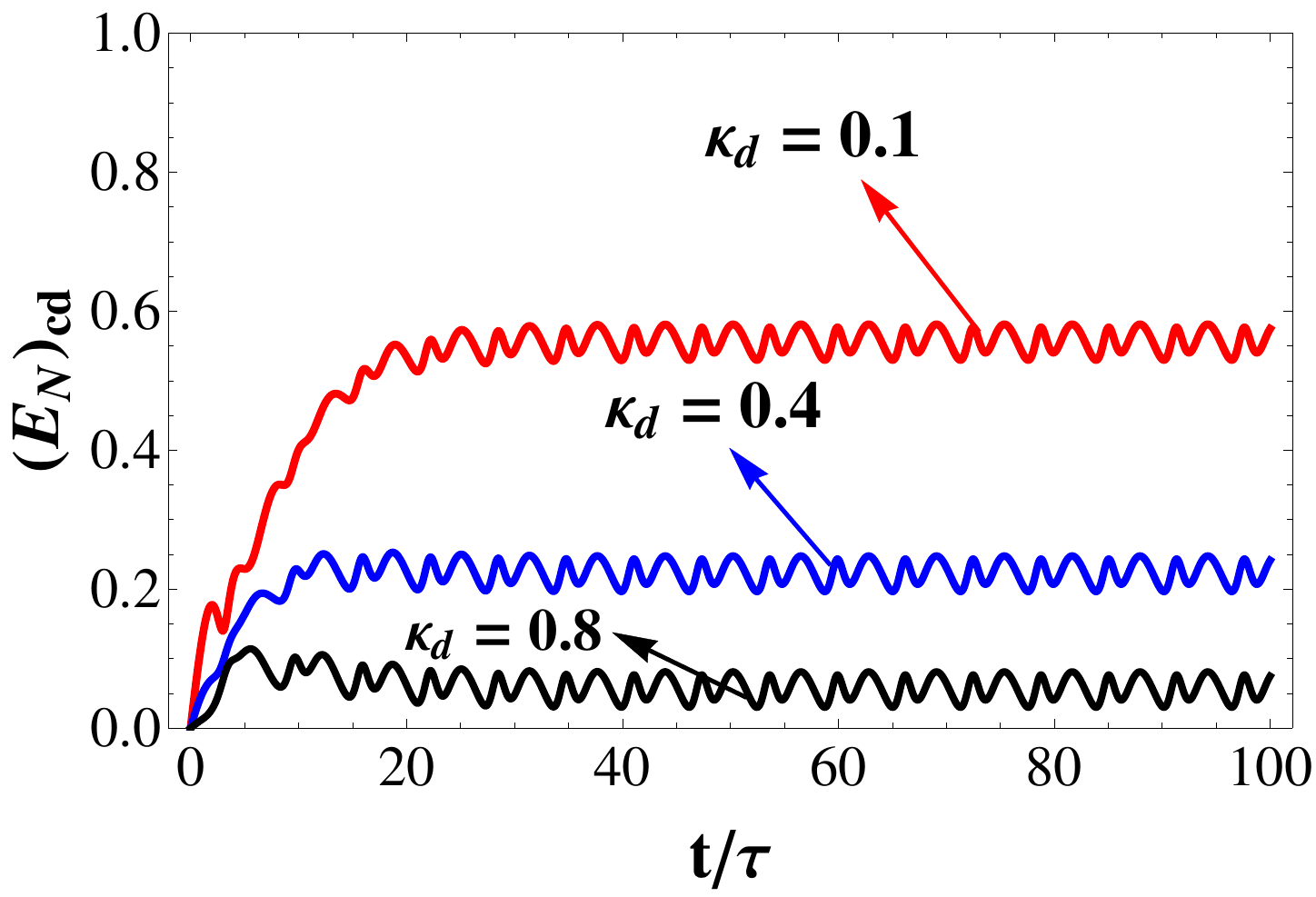} \\
					\end{tabular}
		\caption{(Color online) Dynamics of the QD-cavity entanglement $(E_{N})_{cd}$ as a function of time for different values of QD decay rate ($\kappa_{d}$). The chosen parameters in units of $\omega_{m}$ are:  $\kappa_{a}$=0.1, $\Delta_{0}$=1 $\Delta_{c}$=1, $g_{0}$=0.3, $G$=0.005, $n_{b}=2$, $\gamma$=0.01, N=1, $\epsilon$=0.6, $E_{0}$= 1. }
	\end{figure}
	
	 \begin{figure}[h]
  	\centering
  	\begin{tabular}{@{}cccc@{}}
  		\includegraphics[width=.45\textwidth]{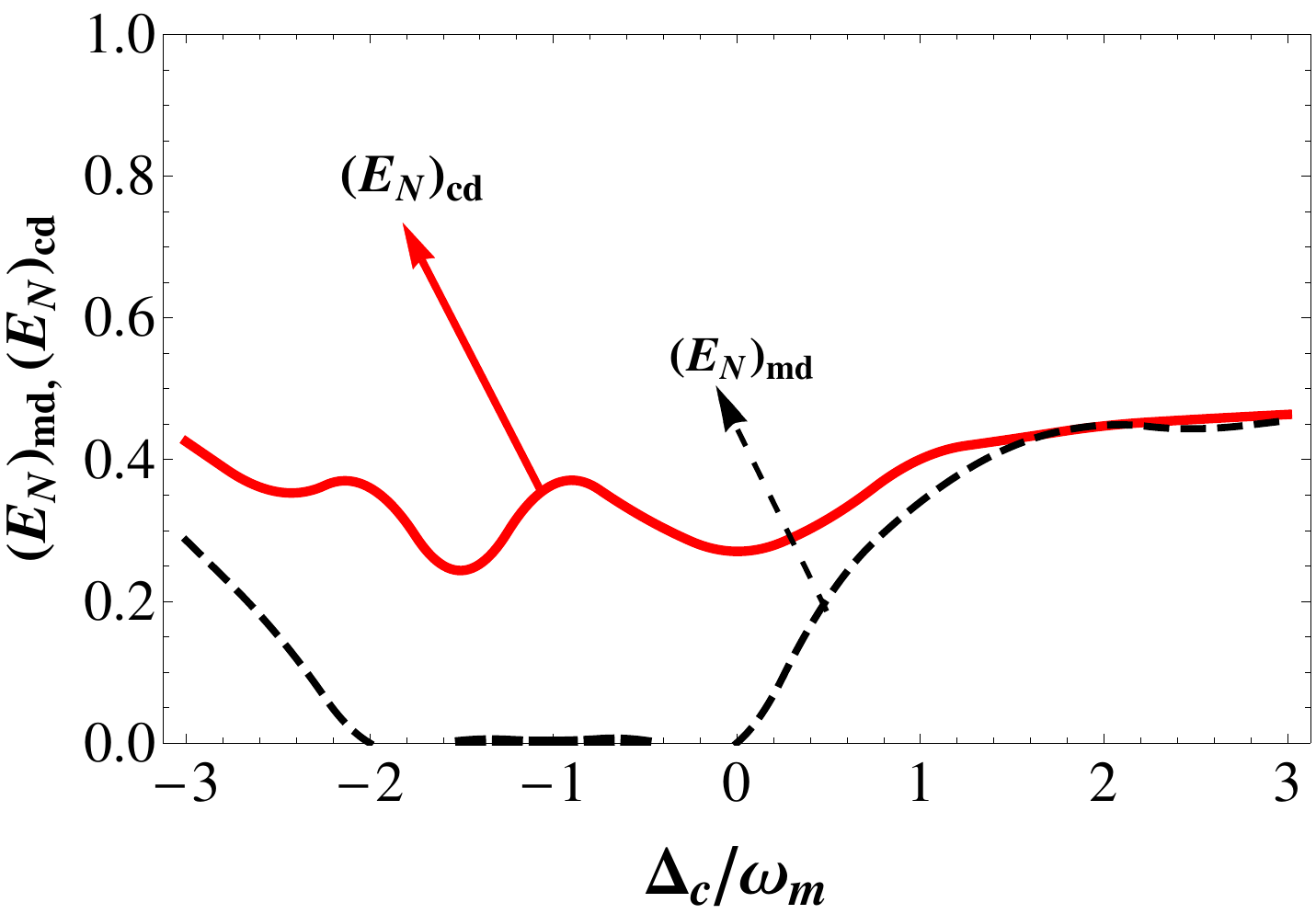} \\
  		  	\end{tabular}
  	\caption{(Color online) Dynamics of the QD- mirror entanglement $(E_{N})_{md}$ and QD-cavity entanglement $(E_{N})_{cd}$ as a function of cavity detuning ($\Delta_{c}$). The chosen parameters in units of $\omega_{m}$ are:  $\kappa_{a}$=0.1, $\kappa_{d}$=0.2, $\Delta_{0}$=1 $\Delta_{c}$=1, $g_{0}$=0.3, $G$=0.005, $\Omega$=1, $n_{b}=2$, $\gamma$=0.01, N=1, $\epsilon$=0.6, $E_{0}$= 1. }
  \end{figure}
	
	In Fig. 7, we plot the time evolution of the QD-mirror entanglement $(E_{N})_{md}$ for two different values of $\Omega$. The dynamical continuous variable entanglement between the QD and the mechanical resonator is generated by their common interaction with the cavity mode.  We notice that as we increase the modulation frequency, initially there is a significant enhancement in entanglement $(E_{N})_{md}$ and then it tends to an average steady-state value around which it oscillates (dynamical entanglement). This enhancement becomes more profound as we increase the modulation frequency ($\Omega=4 \omega_{m}$). Moreover, in large time limit, we note that entanglement acquires fast periodic modulation around a steady value as we increase the modulation frequency (see plot, 7(a) and 7(b)). The average steady value for $\Omega$=4$\omega_{m}$ is significantly higher (=0.8) compared to that for $\Omega$=$\omega_{m}$ (=0.45). The maximum entanglement attained for $\Omega$=$\omega_{m}$ is $(E_{N})_{md}$= 0.52 while for $\Omega$=4$\omega_{m}$ it is $(E_{N})_{md}$=0.95. For $\Omega$=4$\omega_{m}$, the amplitude of oscillations are also much larger compared to that for $\Omega$=$\omega_{m}$. This indicates the possibility of dynamic entanglement between the QD and the mechanical resonator mode in the absence of a direct coupling between them, enabling transfer of information between the QD and the movable mirror. In future quantum internet technologies, such systems can enable quantum communication between a solid state qubit and a mechanical mode via the optical mode. It is then obvious that the optomechanical coupling strength $G$ and the coupling strength $g_{0}$ between the QD and the cavity mode are two suitable parameters to tune the entanglement between the QD and the mechanical oscillator. In Eqns.(6)-(9), we eliminate the time dependence using the technique of few-mode expansion \citep{zhang} and obtain the steady-state solution for $q_{s}=\left < q \right >$ as

	\begin{equation}
q_{s}= \frac{\chi |\sigma_{-}|^{2}}{g_{0}^{2}N^{2}}\left[ \frac{\Delta_{0}^{2}N^{2}}{4} + \kappa_{d}^{2} \right] + \chi C,
\end{equation}

where, $\chi$= G/$\omega_{m}$ and C= $\frac{\epsilon^{2}}{2\Omega^{2}}$.

Eqn.(33) demonstrates the indirect correlation between the mechanical resonator and the QD. It is obvious that the steady mechanical displacement decreases as $g_{0}$ increases and this is illustrated in Fig.8(a), where we show the variation of $q_{s}$ as a function of $g_{0}$. In Fig.8(b), we show the time evolution of $(E_{N})_{md}$ for three different values of $g_{0}$. As expected, the entanglement decreases as $g_{0}$ increases since from Eqn.(33) and Fig.8(a), $q_{s}$ decreases as $g_{0}$ increases.
	
	In Fig.9, we show the dynamics of the QD-cavity entanglement $(E_{N})_{cd}$ as a function of time for different values of QD decay rate ($\kappa_{d}$). It is obvious from the Fig.9 that a fast decaying QD generates less entanglement. A large decoherence of QD, cavity or the mechanical oscillator  has a detrimental effect on the entanglement dynamics. It is thus desirable to work with a good Q-factor cavity and mechanical resonator along with a QD with low decay rate controlled by an electric bias.

	\begin{figure}[htp]
		\centering
		\begin{tabular}{@{}cccc@{}}
			\includegraphics[width=.45\textwidth]{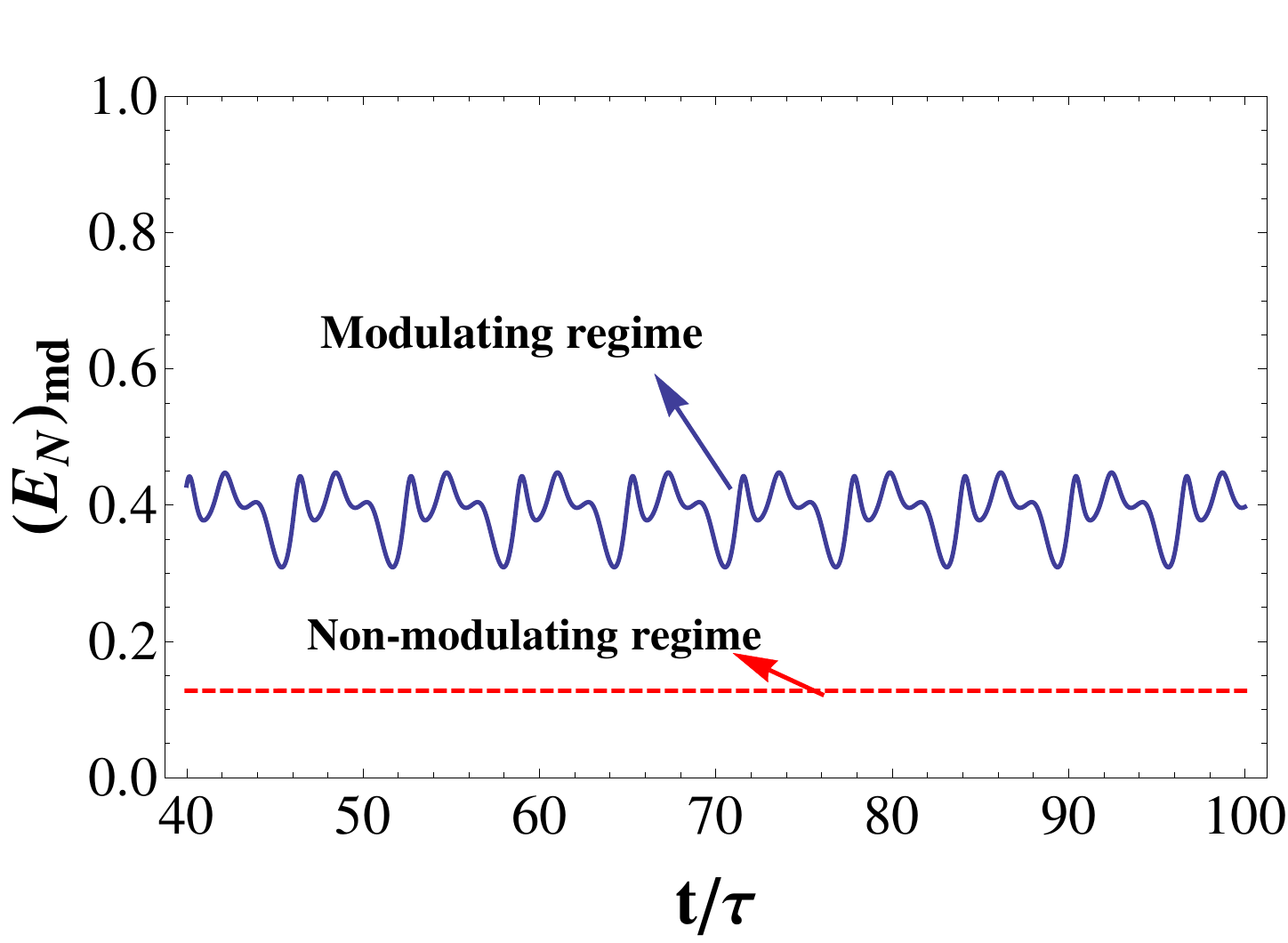}\\
		\end{tabular}
		\caption{(Color online) Dynamics of the QD-mirror entanglement $(E_{N})_{md}$ in the long time limit with modulation (solid line) and without modulation (dashed line). Other parameters in units of $\omega_{m}$ are:  $\kappa_{a}$=0.1, $\kappa_{d}$=0.2, $\Delta_{0}$=1 $\Delta_{c}$=1, $g_{0}$=0.3, $G$=0.005, $\Omega$=1, $n_{b}=2$, $\gamma$=0.01, N=1, $\epsilon$=0.6, $E_{0}$= 1. }
	\end{figure}

	\begin{figure}[htp]
		\centering
		\begin{tabular}{@{}c@{}}
			\includegraphics [scale=0.45]{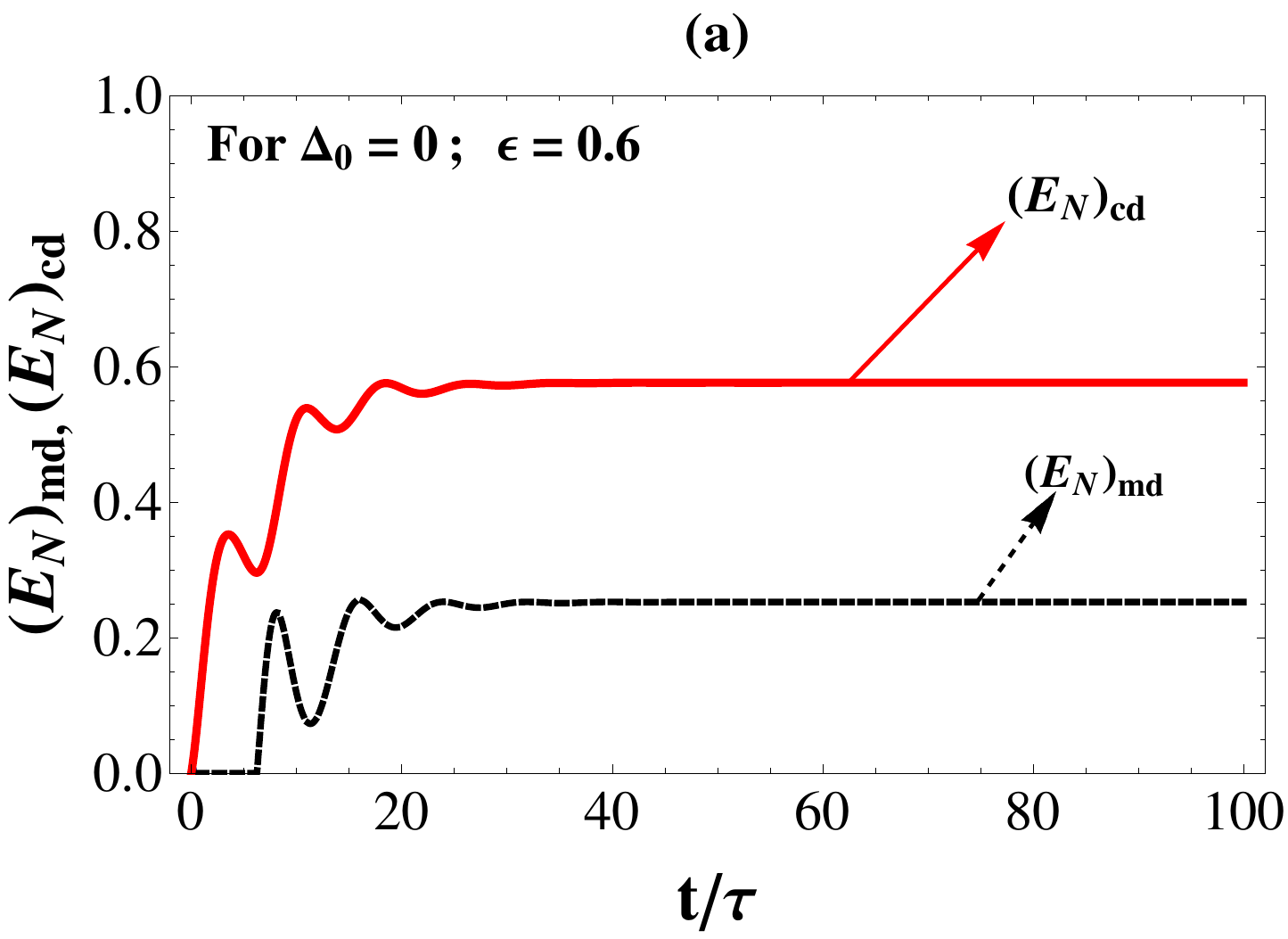}  \includegraphics [scale=0.45] {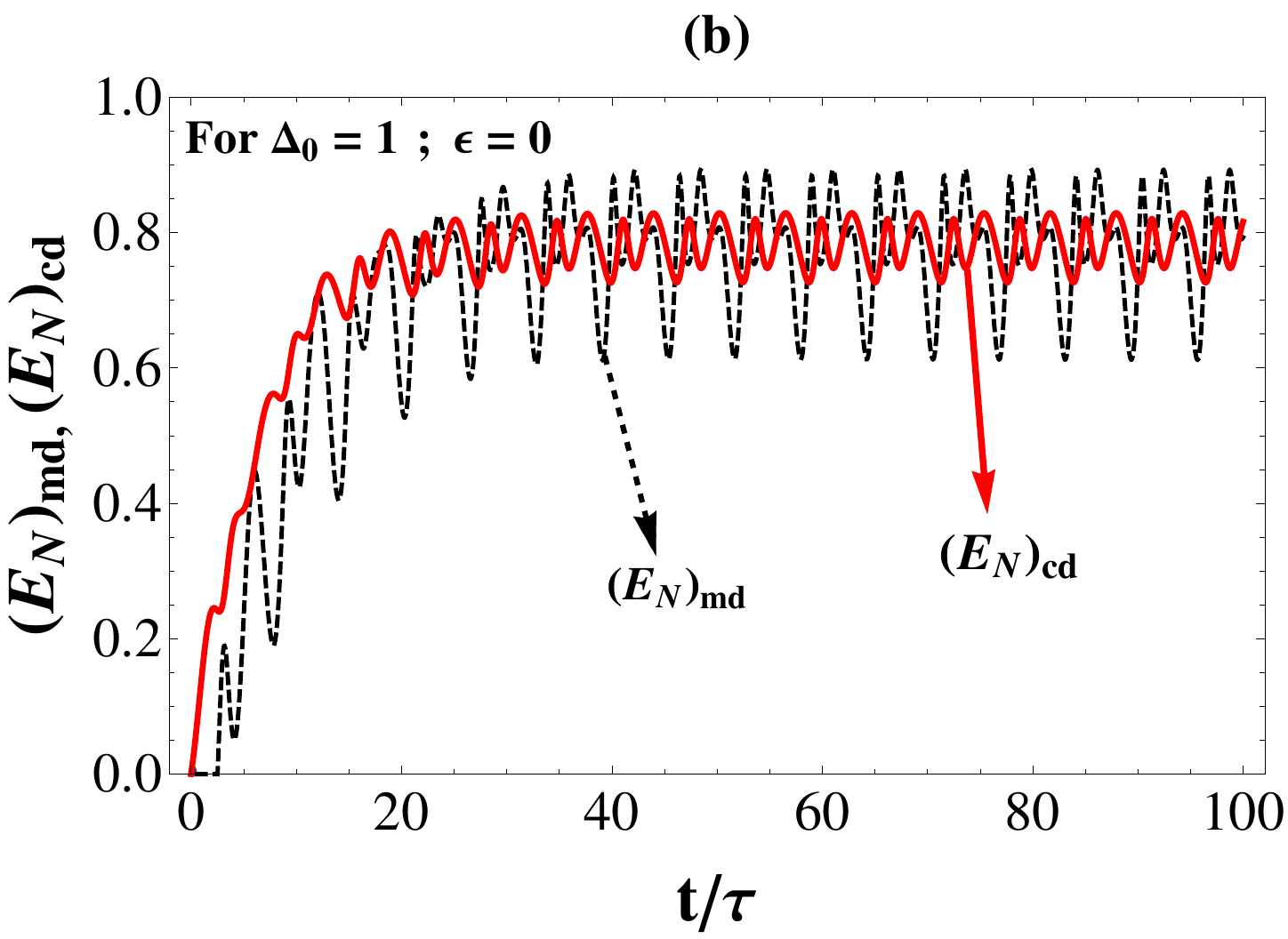}\\
		\end{tabular}
		\caption{(Color online) Dynamics of the QD- mirror entanglement $(E_{N})_{cd}$ and QD-cavity entanglement $(E_{N})_{md}$ as a function of time. The chosen parameters in units of $\omega_{m}$ are: (a)- $\Delta_{0}$=0, $\epsilon$=0.6. (b)- $\Delta_{0}$=1, $\epsilon$=0. Rest of the parameters are: $\kappa_{a}$=0.1, $\kappa_{d}$=0.2, $\Delta_{c}$=1, $g_{0}$=0.3, $G$=0.005, $\Omega$=1, $n_{b}=2$, $\gamma$=0.01, N=1, $E_{0}$= 1.}
	\end{figure}

	Fig.10 shows the plot of QD- mirror entanglement $(E_{N})_{md}$ and QD-cavity entanglement $(E_{N})_{cd}$ as a function of cavity detuning ($\Delta_{c}$). Both the values of entanglement approach the same value as the detuning increases towards large positive values. We note that both $(E_{N})_{cd}$ and $(E_{N})_{md}$ show decreasing behavior between $\Delta_{c}$=0 to $\Delta_{c}$=-2. In particular in this $\Delta_{c}$ window $(E_{N})_{md}$ completely vanishes. Since the coupling between the QD and mirror is indirect, we see that the entanglement  $(E_{N})_{md}$ is less than $(E_{N})_{cd}$. In Fig.11, we plot $(E_{N})_{md}$ for two regimes: with and without modulation. Interestingly the modulation generates a large amount of dynamical entanglement compared to the non-modulating case which generates a low stationary entanglement. This can be explained by the fact that as noted previously, modulation enhances the energy exchange between the various modes.  Fig.12 depicts the entanglement dynamics in the presence of only one modulation. In Fig.12(a), $\Delta_{0}=0$ and $\epsilon=0.6$ (i.e. the modulation of the QD is switched off and only the amplitude of the pump laser is modulated). Both the entanglement measures $(E_{N})_{md}$ and $(E_{N})_{cd}$ approaches a stationary entanglement at about $40 \tau$. This time of $40 \tau$ interestingly is the same time around which the phase space trajectories of the mean values of the cavity mode and the mechanical oscillator's position and momentum approaches the limit cycle. As noted earlier, the fluctuation energy of the cavity and exciton modes also attain an average steady value at about $40 \tau$. In Fig.12(b), $\Delta_{0}=1$, $\epsilon=0$ (ie. only the energy level of the QD is modulated). Interestingly even though the entanglement never attains a stationary value but it does exhibit a higher value compared to that found in Fig.12(a). Thus modulation of the QD is a tradeoff between higher value of entanglement and stationary value. A comparison of Figs.(11) and (12) tells us that modulating the QD frequency generates a larger value of entanglement but introducing the second modulation on the input laser intensity leads to a decrease in the entanglement.
	
	Finally, we can discuss briefly the procedure of experimentally detecting the entanglement that we have discussed in this work. For the optical cavity field, homodyning the cavity output using a local oscillator with an appropriate phase can be utilized to measure the quadratures directly \citep{vitali_last}. To measure the mechanical quadratures of the oscillating mirror, one can utilize a second cavity to measure the position and momentum of the oscillator \citep{vitali_last}.
	
	\section{Conclusion}
	
	In summary, we have introduced and analyzed the quantum dynamics of a solid-state based optomechanical device, consisting of a single quantum dot whose frequency and the amplitude of the external pump laser is periodically modulated. We focus on the evolution of the system in the long time limit since the mean values of the system variables acquire the same periodicity of the modulation in this time limit. We demonstrate that eventhough there is no direct coupling between the QD and the mechanical oscillator, we are still able to control the transfer of exciton fluctuation energy to the mechanical oscillator mode indirectly via the cavity mode. We further show that a high degree of entanglement between the QD and the mechanical oscillator (also QD and cavity mode) can be achieved and controlled. Finally, we show that modulating only the QD frequency leads to a much higher degree of entanglement compared to when either the pump intensity is only modulated or when both the modulations are applied simultaneously. This proposal is particularly attactive since solid-state based systems can easily be fabricated and integrated into large networks. This study shows that in the future this sort of hybrid semiconductor-optomechanical device can be used for data signal transfer, storage and can become part of a broader quantum information processing unit.

	\begin{acknowledgements}
		\textbf{P.K Jha}, \textbf{Aranya B. Bhattacherjee} and \textbf{Vijay Bhatt} are thankful to \textbf{Department of Science and Technology DST(SERB), Project No. EMR/2017/001980, New Delhi} for the financial support. \textbf{Aranya B. Bhattacherjee} and \textbf{Souri Banerjee} is grateful to \textbf{BITS Pilani, Hyderabad campus} for the facilities to carry out this research.
	\end{acknowledgements}

	\vspace*{1cm}
	
	\section{Appendix A}
	
	Gaussian states are of central importance in the context of the continuous-variable (CV) quantum information. These are the states with Gaussian Wigner function and are characterized as well as solved entirely by the field quadrature operators' first and second moments \citep{bra}. 
	Further, to discuss entanglement in CV systems, we consider a very prototypical CV entangled state, i.e., a two-mode Gaussian state. The following covariance matrix can represent this type of state,
	
	\begin{equation}
	\tag{A1}
	V_{2}=
	\begin{bmatrix}
	A  & C \\
	C^{T} & B \\
	\end{bmatrix},
	\end{equation}

	where A,B and C are 2 $\times$ 2 block matrices which descirbe the local properties mode A, mode B and the intermode correlation between A and B. The continuous variable entanglement between the two modes can be well calculated by using the logarithmic negativity $E_{N}$ \citep{vidal,adesso} as,
	
	\begin{equation}
	\tag{A2}
	E_{N}= max[0,-ln 2\nu^{-}].
	\end{equation}
	Where $\nu^{-}$ = $\left(\Sigma(V_{2})-\sqrt{\Sigma(V_{2})^{2}-4 det V_{2}}\right)^{1/2}2^{-1/2}$ is the smallest symplectic eigenvalue of the partial transpose of V2 with $\Sigma(V_{2})$= det(A) + det(B) - 2det(C). The Gaussian state is said to be entangled ($E_{N}$) if it follow the Simon's necessary and sufficient nonpositive partial transpose criteria \citep{simon}.
	
\end{document}